\documentclass[sigconf,table]{acmart}
\usepackage{acronym}
\usepackage[ruled,linesnumbered]{algorithm2e}
\usepackage{balance} 
\usepackage{multirow}
\usepackage{pifont}
\usepackage{siunitx}
\usepackage{subcaption}
\usepackage{textcomp}

\newcommand{\SanctionedAddressCount}{$170$\xspace}
\newcommand{\InitSanctionedBalanceETH}{$499{,}769.74$ ETH\xspace}
\newcommand{\InitSanctionedBalanceUSD}{$983$ million USD\xspace}
\newcommand{\EndSanctionedBalanceETH}{$145{,}579.38$ ETH\xspace}
\newcommand{\EndSanctionedBalanceUSD}{$286$ million USD\xspace}
\newcommand{\EndStudyTime}{March 21, 2025\xspace}
\newcommand{\InitTCPercentage}{$44.17\%$\xspace}
\newcommand{\EndTCPercentage}{$99.12\%$\xspace}

\newcommand{\StudyBlockCount}{$6.79$ million\xspace}

\newcommand{\TCLaunderedMoneyDuringSanctionUSD}{$2$ billion USD\xspace}
\newcommand{\TCEstablishDate}{December 16, 2019\xspace}
\newcommand{\TCAddSanctionDate}{August 8, 2022\xspace}
\newcommand{\TCAddSanctionBlock}{Block \href{https://etherscan.io/block/15302392}{15302392}\xspace}
\newcommand{\TCRemoveSanctionDate}{March 21, 2025\xspace}
\newcommand{\TCRemoveSanctionBlock}{Block \href{https://etherscan.io/block/22097863}{22097863}\xspace}
\newcommand{\TCPreSanctionPeriod}{$966$ days\xspace}
\newcommand{\TCSanctionPeriod}{$957$ days\xspace}

\newcommand{\TCDepositPreSanTxCount}{$150{,}403$\xspace}
\newcommand{\TCDepositPostSanTxCount}{$42{,}753$\xspace}
\newcommand{\TCDepositTxCountDecreasePercentage}{$71.57\%$\xspace}
\newcommand{\TCWithdrawPreSanTxCount}{$139{,}404$\xspace}
\newcommand{\TCWithdrawPostSanTxCount}{$44{,}283$\xspace}
\newcommand{\TCWithdrawTxCountDecreasePercentage}{$68.23\%$\xspace}

\newcommand{\TCDepositPreSanAddrCount}{$39{,}397$\xspace}
\newcommand{\TCDepositPostSanAddrCount}{$9{,}175$\xspace}
\newcommand{\TCDepositAddrCountDecreasePercentage}{$76.71\%$\xspace}
\newcommand{\TCWithdrawPreSanAddrCount}{$58{,}567$\xspace}
\newcommand{\TCWithdrawPostSanAddrCount}{$20{,}734$\xspace}
\newcommand{\TCWithdrawAddrCountDecreasePercentage}{$64.60\%$\xspace}

\newcommand{\TCDepositPreSanVolume}{$3.49M$\xspace}
\newcommand{\TCDepositPostSanVolume}{$1.01M$\xspace}
\newcommand{\TCDepositVolumeDecreasePercentage}{$71.03\%$\xspace}
\newcommand{\TCWithdrawPreSanVolume}{$3.27M$\xspace}
\newcommand{\TCWithdrawPostSanVolume}{$1.08M$\xspace}
\newcommand{\TCWithdrawVolumeDecreasePercentage}{$66.80\%$\xspace}

\newcommand{\DustAttackUsedPreciseETH}{$12.00$ ETH\xspace}
\newcommand{\DustAttackTaintedETH}{$9$ million ETH\xspace}
\newcommand{\DustAttackTaintedPreciseETH}{$9{,}034{,}251.58$ ETH\xspace}
\newcommand{\DustAttackBlockCount}{$14{,}400$\xspace}
\newcommand{\DustAttackTime}{$48$ hours\xspace}
\newcommand{\DustAttackAmplify}{$750{,}000 \times$\xspace}

\newcommand{\AttackerUseTCPercentage}{$78.33\%$\xspace}
\newcommand{\EthereumIncidentCount}{$60$\xspace}

\acrodef{OFAC}[OFAC]{Office of Foreign Assets Control}
\acrodef{FATF}[FATF]{Financial Action Task Force}
\acrodef{TC}[TC]{Tornado Cash}
\acrodef{NFT}[NFT]{Non-Fungible Token}
\acrodef{AML}[AML]{Anti-Money Laundering}
\acrodef{CTF}[CTF]{Counter-Terrorism Financing}
\acrodef{KYC}[KYC]{Know-Your-Customer}
\acrodef{KYT}[KYT]{Know-Your-Transaction}
\acrodef{DeFi}[DeFi]{Decentralized Finance}
\acrodef{ZKP}[ZKP]{Zero-Knowledge Proof}
\acrodef{P2P}[P2P]{Peer-to-Peer}
\acrodef{EVM}[EVM]{Ethereum Virtual Machine}
\acrodef{UTXO}[UTXO]{Unspent Transaction Output}
\acrodef{MEV}[MEV]{Maximal Extractable Value}
\acrodef{CEX}[CEX]{Centralized Exchange}
\acrodef{DEX}[DEX]{Decentralized Exchange}
\acrodef{AMM}[AMM]{Automated Market Maker}
\acrodef{RPC}[RPC]{Remote Procedure Call}
\acrodef{CLT}[CLT]{Command Line Tool}

\newcommand{\circledP}{\raisebox{.5pt}{\textcircled{\raisebox{-.9pt}{P}}}}
\newcommand{\circledOne}{\raisebox{.5pt}{\textcircled{\raisebox{-.9pt}{1}}}}
\newcommand{\circledTwo}{\raisebox{.5pt}{\textcircled{\raisebox{-.9pt}{2}}}}
\newcommand{\circledThree}{\raisebox{.5pt}{\textcircled{\raisebox{-.9pt}{3}}}}

\usepackage{enumitem}
\setlist{
    nosep,
    leftmargin=*,
    align=left,
}

\usepackage{textcomp}
\DeclareFontFamily{OMS}{LinuxLibertineT-TLF}{}
\DeclareFontShape{OMS}{LinuxLibertineT-TLF}{m}{n}{<-> ssub * cmsy/m/n}{}

\definecolor{mylightyellow}{HTML}{FCE1A4}

\AtBeginDocument{%
  }

\copyrightyear{2026}
\acmYear{2026}
\setcopyright{cc}
\setcctype{by-nc-nd}
\acmDOI{10.1145/3774904.3792715}
\acmConference[WWW '26]{Proceedings of the ACM Web Conference 2026}{April 13--17, 2026}{Dubai, United Arab Emirates}
\acmBooktitle{Proceedings of the ACM Web Conference 2026 (WWW '26), April 13--17, 2026, Dubai, United Arab Emirates}
\acmISBN{979-8-4007-2307-0/2026/04}

\begin{document}

\title{Evasion Under Blockchain Sanctions}

\author{Endong Liu}
\orcid{0009-0004-7574-4001}
\affiliation{
  \institution{University of Birmingham}
  \city{Birmingham}
  \country{United Kingdom}
}
\email{exl309@student.bham.ac.uk}

\author{Mark Ryan}
\orcid{0000-0002-1632-497X}
\affiliation{
  \institution{University of Birmingham}
  \city{Birmingham}
  \country{United Kingdom}
}
\email{m.d.ryan@bham.ac.uk}

\author{Liyi Zhou}
\authornote{Co-supervised this research project.}
\orcid{0000-0002-2820-9872}
\affiliation{
  \institution{The University of Sydney}
  \city{Sydney}
  \country{Australia}
}
\email{liyi.zhou@sydney.edu.au}

\author{Pascal Berrang}
\authornotemark[1]
\orcid{0000-0002-9194-9603}
\affiliation{
  \institution{University of Birmingham}
  \city{Birmingham}
  \country{United Kingdom}
}
\email{p.p.berrang@bham.ac.uk}

\renewcommand{\shortauthors}{Endong Liu, Mark Ryan, Liyi Zhou, and Pascal Berrang}

\begin{abstract}
Sanctioning blockchain addresses has become a common regulatory response to malicious activities. However, enforcement on permissionless blockchains remains challenging due to complex transaction flows and sophisticated fund-obfuscation techniques. 

Using cryptocurrency mixing tool Tornado Cash as a case study, we quantitatively assess the effectiveness of U.S. Office of Foreign Assets Control (OFAC) sanctions over a 957-day period, covering 6.79 million Ethereum blocks and 1.07 billion transactions. Our analysis reveals that while OFAC sanctions reduced overall Tornado Cash deposit volume by 71.03\% to approximately 2 billion USD, attackers still relied on Tornado Cash in 78.33\% of Ethereum-related security incidents, underscoring persistent evasion strategies. 

In this paper, we identify three significant, structural limitations in current sanction enforcement practices: \textit{(i)} fragmented censorship in blockchain consensus and application layer; \textit{(ii)} the complexity of obfuscation virtual asset services exploited by users; and \textit{(iii)}~the susceptibility of na{\"i}ve binary sanction classifications to dusting attacks. Our analysis and findings contribute to ongoing discussions around regulatory effectiveness in Decentralized Finance by providing empirical evidence, clarifying enforcement challenges, and informing future compliance strategies in response to sanctions and blockchain-based security risks.
\end{abstract}

\ccsdesc[500]{Security and privacy}
\keywords{Blockchain; Sanction; Compliance; Ethereum; Tornado Cash; Transaction Tracking; Fund Obfuscation; Decentralized Finance; DeFi}

\maketitle

\section{Introduction}
\label{sec:introduction}

Sanctioning or blacklisting blockchain addresses is a strategy used by regulators to counter activities such as money laundering and terrorist financing. Despite ongoing efforts, enforcing sanctions on permissionless blockchains remains challenging, due to the complexity of transaction flows and the ease with which assets can be moved across addresses and protocols. For example, the U.S. \acf{OFAC} sanctioned~\SanctionedAddressCount addresses on Ethereum~\cite{chainalysis_sanctions_screening} that collectively held an initial balance of~\InitSanctionedBalanceETH (\InitSanctionedBalanceUSD) at the time of designation. Following these sanctions, adversaries transferred funds to newly created addresses and employed a range of obfuscation techniques to evade detection. By the end of the observation period (\EndStudyTime), only~\EndSanctionedBalanceETH (\EndSanctionedBalanceUSD) remained.

Among these \ac{OFAC}-sanctioned addresses, \ac{TC} represents the most significant case, constituting~\InitTCPercentage of the initial and~\EndTCPercentage of the remaining balance. As a privacy-focused cryptocurrency mixing pool, \ac{TC} can break the on-chain linkability between deposit and withdrawal transactions. From our empirical analysis, \ac{TC} processed over \TCLaunderedMoneyDuringSanctionUSD during the sanctioned period from \TCAddSanctionDate~\cite{treasury2022tornadocash} to \TCRemoveSanctionDate~\cite{treasury2025tornadocash} (\TCSanctionPeriod). Remarkably, in~\AttackerUseTCPercentage of \EthereumIncidentCount Ethereum-related security incidents during this period, attackers continued to leverage \ac{TC}, illustrating persistent circumvention of sanctions by sophisticated actors. 

While prior research has predominantly focused on tracing security incidents~\cite{DBLP:conf/ccs/GomezMC22, DBLP:journals/tdsc/WahrstatterGKS23, DBLP:journals/tifs/WuLFYCZS24, DBLP:conf/www/0007WYFZY24, DBLP:conf/sp/ZhouXECWWQWSG23} or reconstructing the linkability of anonymous transactions~\cite{DBLP:conf/www/WangCQZGBLG23, DBLP:conf/www/KovacsS24, DBLP:journals/tdsc/WahrstatterTS24, DBLP:journals/tifs/DuCSZH24}, our study broadly assesses regulatory effectiveness by systematically evaluating the impact of sanctions. We identify three structural limitations in the current sanction enforcement framework as our main contributions: 

\begin{itemize}
    \item \textbf{Fragile Sanction Enforcement:} 
    We empirically evaluate the effectiveness of \ac{OFAC} sanctions on Ethereum ($\S$\ref{sec:sanction_effectiveness}) across a comprehensive dataset spanning 957 days of the entire sanction period, encompassing 6.79 million blocks and 1.07 billion transactions. We demystify the censorship gaps in consensus layer ($\S$\ref{subsec:consensus_gap}) and application layer ($\S$\ref{subsec:application_gap}), and provide temporal analysis on user reactions ($\S$\ref{subsec:user_reaction}). Our findings indicate that, although the overall \ac{TC} deposit volume decreased by \TCDepositVolumeDecreasePercentage during the sanction period compared to a similar pre-sanction timeframe, \ac{TC} remains the primary tool for laundering malicious proceeds in Ethereum-related security incidents. 

    \item \textbf{Comprehensive Obfuscation Patterns and Ecosystem:} We detect notable user strategies ($\S$\ref{subsec:behavioral_pattern}), such as short dormant time (used to reduce the risk of assets being frozen), split-and-merge patterns (associated with money laundering), and test-depositing transactions (used to probe sanction enforcement policies). We further find that users can leverage the entire Ethereum service landscape to obscure fund origins ($\S$\ref{subsec:usage_of_service_providers}), including cross-chain bridges, \acp{DEX}, \acp{CEX}, alternative privacy-enhancing tools, and other \ac{DeFi} platforms.

    \item \textbf{Na{\"i}ve Binary Classification Vulnerability:} We provide an argument to demonstrate that binary classification systems (sanctioned / non-sanctioned) used in the \ac{OFAC} list exhibit inherent vulnerabilities to \emph{dusting attacks} in Appendix~\ref{appendix:binary_impossibility}, where small amounts of sanctioned funds are strategically distributed to arbitrary addresses, finally resulting denial of service. Our analysis also reveals that \DustAttackUsedPreciseETH from sanctioned services were used to dust addresses controlling at least \DustAttackTaintedETH (over \DustAttackAmplify) within \DustAttackTime after sanctions took effect ($\S$\ref{subsubsec:dusting}).
\end{itemize}


\section{Background}
\label{sec:background}

This section outlines the necessary background for the paper.

\subsection{Ethereum and Its Service Ecosystem}

Ethereum is a permissionless blockchain system that supports smart contracts, which are programs that can be deployed and executed by anyone on the network. Unlike Bitcoin's \ac{UTXO} model~\cite{nakamoto2008bitcoin}, Ethereum employs an account-based model where each address maintains a state including its balance and other attributes~\cite{wood2025ethereum}. The native cryptocurrency, Ether ($1$ ETH $= 10^{18}$ Wei), serves both as a medium of exchange and as ``gas'' to pay for all on-chain computational operations. Transactions are processed by miners (Proof-of-Work) or validators (Proof-of-Stake after the ``Merge'' in September 2022), who receive rewards for including transactions. Its ecosystem encompasses various services:

\textbf{Privacy-Enhancing Tools}: Mixing services like \ac{TC} aim to break the on-chain linkability between deposit and withdrawal transactions. It operates via fixed-denomination mixing pools (e.g., 0.1 ETH, 1 ETH, 10 ETH, and 100 ETH), utilizing \acp{ZKP} to provide a cryptographic proof of commitment ownership without revealing specific commitments. Other tools like Umbra~\cite{umbra_app} and Railgun~\cite{railgunwiki} leverage stealth addresses to prevent observers from connecting multiple payments to the same entity. While these tools serve legitimate privacy needs, they fundamentally challenge traditional regulatory approaches.

\textbf{Exchanges}: There are two primary types of cryptocurrency exchanges. \acp{DEX} like Uniswap~\cite{uniswap_protocol} and 1inch~\cite{1inch_protocol} operate entirely on-chain through smart contracts, enabling peer-to-peer trading without custodial intermediaries. Most operate through an \ac{AMM}, which use mathematical formulas to price assets based on the ratio of tokens in liquidity pools. \acp{CEX} such as Binance~\cite{binance_exchange} and Coinbase~\cite{coinbase_exchange} function as traditional financial intermediaries, maintaining order books, and custodying user funds. \acp{CEX} act as gateways between fiat and crypto, typically enforcing different levels of compliance requirements.

\textbf{Cross-Chain Bridges}: Services that facilitate asset transfers between different blockchain networks to expand interoperability. Unlike token standards (e.g., ERC-20~\cite{erc20_standard}) which provide consistent interfaces, cross-chain bridges lack standardization in their implementation, message formats, and identifier systems. The differences in their architectures~\cite{DBLP:conf/sp/AugustoBCVZH24} create significant technical challenges for cross-chain transaction monitoring and fund tracing.

\textbf{Other \ac{DeFi} Services}: Applications in \ac{DeFi} providing lending, \ac{NFT} trading, and other financial services through smart contracts without traditional intermediaries~\cite{DBLP:conf/aft/Werner0GKHK22}.

\subsection{Regulatory and Enforcement}

The pseudonymous and decentralized nature of blockchains challenges traditional censorship frameworks.

\textbf{Regulatory Evolution}: Traditional \ac{AML} and \ac{CTF} frameworks require \acf{KYC} for customer identification and \acf{KYT} for transaction monitoring. The \acf{FATF} has extended these requirements to virtual asset service providers through recommendations like the ``Travel Rule''~\cite{fatf2023update}, which conflicts with blockchain's pseudonymous design.

\textbf{Blockchain Sanctions}:  The U.S. \ac{OFAC} has expanded their sanctions regimes to include specific blockchain addresses and services. Sanctions on \ac{TC} in August 2022 targeted an autonomous smart contract system that is abused to launder over 7 billion USD worth of cryptocurrency, including over 455 million USD stolen by the North Korean state-sponsored Lazarus Group~\cite{treasury2022tornadocash}. This action classifies privacy technology itself as a compliance risk, marking a significant escalation in regulatory response to blockchain privacy tools. As a result, recent works~\cite{DBLP:conf/ndss/0007Q024, DBLP:journals/tifs/ZhangXH24, DBLP:journals/popets/BadertscherSW24} and protocols~\cite{railgunwiki, zkbob} consider compliance requirements in their designs and implementations.

\textbf{Enforcement Through Multiple Vectors}: The decentralized nature of blockchain systems has necessitated multi-layered censorship approaches that operate at different levels of the technology stack: \textit{(i)} block producers can refuse to include transactions involving sanctioned addresses~\cite{DBLP:conf/www/WahrstatterEYZQ24}, \textit{(ii)} user interfaces and front-ends of \ac{DeFi} applications can block addresses that interact with sanctioned services, and \textit{(iii)} stablecoin providers and centralized services can freeze assets or close accounts based on their blacklist mechanisms.


\section{Models and Metrics}
\label{sec:model}

This section formalizes our system model, threat model, and quantitative measures we use throughout the evaluation.

\subsection{System Model}

This paper follows the system model in the previous works \cite{DBLP:conf/sp/ZhouXECWWQWSG23,DBLP:conf/www/WahrstatterEYZQ24}: \textbf{Consensus Layer.} A leader is selected to append a block to ledger, who can capture inclusion and sequencing choices that earn rewards and can enact or avoid censorship. \textbf{Application Layer.} Smart contracts maintain state and execute bounded logic; when blacklist or oracle checks are embedded, enforcement becomes binding and auditable on-chain. \textbf{Auxiliary Services.} (as Off-Chain Application Layer) Front-ends and oracles mediate access and data off-chain, adding friction that is bypassable by direct protocol interaction.

To ease the understanding of the following paragraphs, we proceed by introducing the following notations and definitions:

\hspace{1pt}

\begin{description}[leftmargin=!]
    \item[Block] $B_i$ where $i$ corresponds to a block identifier.
    
    \item[Blockchain] $\mathcal{L} = \{B_0, \dots, B_i, \dots\}$ also called \emph{Distributed Ledger}.
    
    \item[Address] $addr \in \textit{ADDR}$, where $\textit{ADDR}$ is the set of valid Ethereum addresses. We use the terms ``address'' and ``account'' interchangeably as each account is identified by a unique address, referring to externally owned accounts and smart contract accounts.

    \item[Sanctioned Address Set] $\textit{ADDR}_\textit{Sanction} \subset \textit{ADDR}$.
    
    \item[Blockchain State] $s \in \mathcal{S}$ includes \textit{(i)} the state of all accounts (e.g., balance and transaction nonce), \textit{(ii)} the state of all smart contracts, and \textit{(iii)} blockchain metadata (e.g., block numbers, timestamps, miner addresses, gas used, gas price, etc.) at a given point in time.

    \item[Operation] $op \in \textit{OP}$, where $\textit{OP}$ is the set of operations that modify the state $s$ (e.g., asset transfers and contract invocations).

    \item[Balance Function] $\mathcal{B}(s, addr) \rightarrow \mathbb{N}_0$ returns the non-negative integer balance (in Wei) for a given address $addr$ in state $s$.

    \item[State Transition] $\mathcal{T}(s, op) \rightarrow s^\prime$ returns the new state $s^\prime$ after the current state $s$ applies operation $op$. Specifically, given a balance operation $op_\mathcal{B}$ that affects addresses $\{addr_{\textit{from}}, addr_{\textit{to}}\} \subset \textit{ADDR}$ and amounts $\{ a_{\textit{sent}}, a_{\textit{received}} \} \subset \mathbb{N}_0$, such that \begin{equation*}
        op_\mathcal{B} \rightarrow \{\, \langle addr_{\textit{from}}, a_{\textit{sent}}\rangle, \langle addr_{\textit{to}}, a_{\textit{received}}\rangle \,\}
    \end{equation*}
    
    the balance state transition $\mathcal{T}_\mathcal{B}$ is \begin{equation*}
    \begin{aligned}
        \mathcal{B}(s^\prime, addr_{\textit{from}}) &= \mathcal{B}(s, addr_{\textit{from}}) - a_{\textit{sent}} \\
        \mathcal{B}(s^\prime, addr_{\textit{to}}) &= \mathcal{B}(s, addr_{\textit{to}}) + a_{\textit{received}}
    \end{aligned}
    \end{equation*}
\end{description}

\subsection{Threat Model}
\label{sec:threat_model}
We consider adversaries who aim to conceal fund flows originating from sanctioned sources. These adversaries may range from individual actors to state-backed organizations (e.g., Lazarus Group). We assume they have full visibility of the blockchain's public data and can undertake the following on-chain actions:

\begin{itemize}
    \item \textbf{Dusting:} Adversaries can send arbitrary amounts to random addresses (including innocent accounts) to create ambiguous connections and inflate the scale of transaction tracing graphs, making it more difficult to pinpoint the true flow of funds.

    \item \textbf{Layering:} Adversaries can generate an unrestricted number of addresses and transactions to split (from one source to multiple destinations) or merge (from multiple sources to one destination) their assets. Furthermore, these addresses can be used only once to undermine straightforward clustering heuristics.

    \item \textbf{Interacting:} Adversaries can create their own smart contracts and can interact with any deployed on-chain service, including swapping tokens on \acp{DEX}, cashing out via \acp{CEX}, bridging to other blockchains, using privacy-enhancing tools, or engaging with \ac{DeFi} lending and \ac{NFT} trading platforms.
\end{itemize}

We assume that adversaries cannot compromise the blockchain's cryptographic foundations, exploit previously unknown smart contract vulnerabilities, or conspire with a majority of block producers or service providers. Our primary goal and focus is to expose the behavioral patterns of adversaries and the ultimate destinations of assets once they leave malicious or sanctioned accounts.

\subsection{Metrics}
\label{subsec:metrics}

In this paper, a more practical tracking algorithm that uses a quantitative impurity metric is applied to precisely measure the propagation and concentration of sanctioned funds throughout the Ethereum ecosystem. We provide the algorithm for updating the corresponding metrics of each affected address as sanctioned assets propagate in Appendix~\ref{appendix:algorithm}.

\hspace{1pt}

\begin{description}[leftmargin=!]
    \item[Impurity Function] $\mathcal{I}(s, addr) \rightarrow \mathbb{N}_0$ returns the non-negative integer amount originated from $\textit{ADDR}_\textit{Sanction}$ in balance (in Wei) for a given address $addr$ in state $s$, such that \begin{equation*}
        0 \leq \mathcal{I}(s, addr) \leq \mathcal{B}(s, addr)
    \end{equation*}

    \item[Impurity Score] $\varphi(s, addr)$ is defined as \begin{equation*}
        \varphi(s, addr) =
        \begin{cases} 
            \frac{\mathcal{I}(s, addr)}{\mathcal{B}(s, addr)}, & \text{if } \mathcal{B}(s, addr) > 0 \\
            0,  & \text{if } \mathcal{B}(s, addr) = 0
        \end{cases}
    \end{equation*}
\end{description}

By definition, $\varphi(s, addr) = 0$ indicates there is no traceable connection between $addr$ and $\textit{ADDR}_\textit{Sanction}$ or no balance under $addr$ at the state $s$, and $\varphi(s, addr) = 1$ shows that all balances under $addr$ are traceable to sanctioned addresses at the state $s$. This score provides a relative measure of an address's historical association with sanctioned outflows. It is important to note that this approach aligns with Ethereum's account model, where the impurity of funds can be treated as an extra state or property of an Ethereum address.


\section{Sanction Effectiveness}
\label{sec:sanction_effectiveness}

To examine the sanction effectiveness, we conduct a comprehensive evaluation using historical Ethereum blockchain data. Our analysis covers the entire \ac{TC} sanction period from \TCAddSanctionBlock ($B_\textit{OFAC}$, \TCAddSanctionDate) to \TCRemoveSanctionBlock ($B_\textit{END}$, \TCRemoveSanctionDate).

\subsection{Gaps in Consensus Layer Censorship}
\label{subsec:consensus_gap}
 
Previous research~\cite{DBLP:conf/www/WahrstatterEYZQ24} showed that block producers can and do filter out transactions associated with \ac{OFAC}-sanctioned entities to avoid including them in their blocks. These censorship behaviors can be driven by external pressures (e.g., regulations) or internal motivations (e.g., ethical or economic considerations). We followed the definition of \emph{Consensus Layer Censorship} in~\cite{DBLP:conf/www/WahrstatterEYZQ24} and further divided it into two different situations as the following definitions:

\begin{definition}[Base Consensus Layer Censorship]
    A blockchain protocol participant intentionally obstructs the inclusion of a transaction where at least one involved address is sanctioned, such that $tx \notin B_i \ | \ B_i \in \mathcal{L}$ and $addr(tx) \cap \textit{ADDR}_\textit{Sanction} \neq \varnothing$.
\end{definition}

However, only a part of blockchain protocol participants actively enforce this basic but strict censorship. Due to the censorship-resistant nature of Ethereum, transactions that move assets from a sanctioned address to other addresses still have a possibility to be included in the distributed ledger. Therefore, we adapt the definition to capture the wider censorship.

\begin{definition}[Derived Consensus Layer Censorship]
    A blockchain protocol participant intentionally obstructs the inclusion of a transaction where all involved addresses are not sanctioned but at least one holds assets that originate from sanctioned addresses at or above a pre-defined threshold, such that $tx \notin B_i \ | \ B_i \in \mathcal{L}$ and $addr(tx) \cap \textit{ADDR}_\textit{Sanction} = \varnothing$ but $\max_{a\in addr(tx)}\varphi(a) \geq \theta$.
\end{definition}

\begin{figure}[tb!]
\centering
\includegraphics[width=\columnwidth]{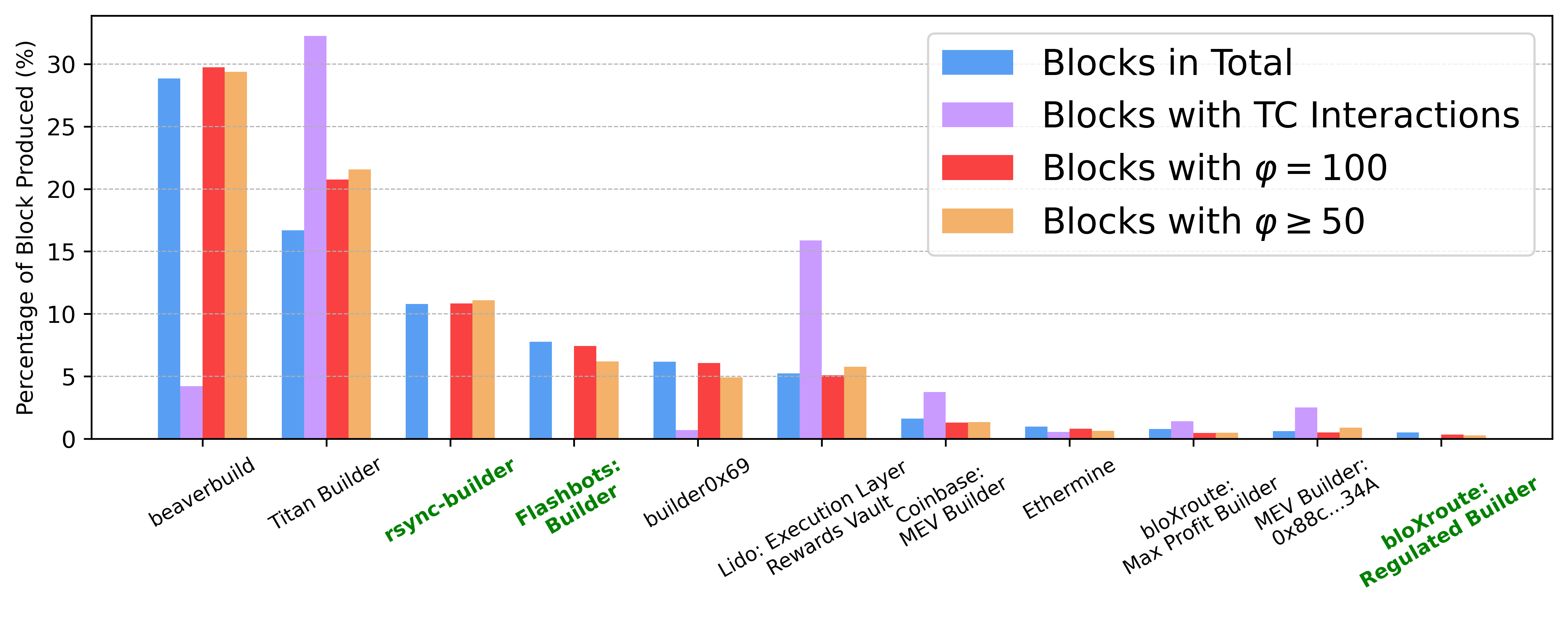}
\caption{Top block producers on Ethereum between $B_{\textit{OFAC}}$ and $B_{\textit{END}}$. Blocks with \ac{TC} interactions (purple) represent blocks that contain direct deposits and withdrawals with four \ac{TC} ETH pools. Blocks with score conditions (red and yellow) indicate blocks that include transactions issued by accounts that meet the condition. Note that \href{https://etherscan.io/address/0x1f9090aae28b8a3dceadf281b0f12828e676c326}{``rsync-builder''}, \href{https://etherscan.io/address/0xDAFEA492D9c6733ae3d56b7Ed1ADB60692c98Bc5}{``Flashbots: Builder''}, and \href{https://etherscan.io/address/0x199D5ED7F45F4eE35960cF22EAde2076e95B253F}{``bloXroute: Regulated Builder''} (in green) follow OFAC sanctions to reject direct \ac{TC} interactions.}
\Description{Top block producers on Ethereum between the OFAC state and the end state. Blocks with TC interactions (purple) represent blocks that contain direct deposits and withdrawals with four TC ETH pools. Blocks with score conditions (red and yellow) indicate blocks that include transactions issued by accounts that meet the condition. Note that ``rsync-builder'', ``Flashbots: Builder'', ``bloXroute: Regulated Builder'' (in green) follow OFAC sanctions to reject direct TC interactions.}
\label{fig:block_producer}
\end{figure}

Our re-evaluation of block producer behaviors using the proposed impurity score $\varphi$ revealed a significant discrepancy between the base censorship and derived handling of downstream assets withdrawn from four \ac{TC} ETH pools. As illustrated in Figure \ref{fig:block_producer}, we visualized top block producers on Ethereum, which covers $80.06\%$ of a \StudyBlockCount block sample between $B_{\textit{OFAC}}$ and $B_\textit{END}$.

Three block producers, \href{https://etherscan.io/address/0x1f9090aae28b8a3dceadf281b0f12828e676c326}{``rsync-builder''}, \href{https://etherscan.io/address/0xDAFEA492D9c6733ae3d56b7Ed1ADB60692c98Bc5}{``Flashbots: Builder''} and \href{https://etherscan.io/address/0x199D5ED7F45F4eE35960cF22EAde2076e95B253F}{``bloXroute: Regulated Builder''}, actively follow \ac{OFAC} sanctions. The \href{https://etherscan.io/address/0x1f9090aae28b8a3dceadf281b0f12828e676c326}{``rsync-builder''} demonstrated the most rigorous censorship. It produced 734,343 blocks (10.81\% of sampled blocks) without including any direct interactions with four \ac{TC} ETH pools. However, our analysis of blocks containing senders with high impurity scores ($\varphi = 100\%$ and $\varphi \geq 50\%$) revealed that \href{https://etherscan.io/address/0x1f9090aae28b8a3dceadf281b0f12828e676c326}{``rsync-builder''} produced these blocks at rates (10.83\% and 11.10\%, respectively) that closely match its overall block production rate (10.81\%). We observed similar patterns with \href{https://etherscan.io/address/0xDAFEA492D9c6733ae3d56b7Ed1ADB60692c98Bc5}{``Flashbots: Builder''} and \href{https://etherscan.io/address/0x199D5ED7F45F4eE35960cF22EAde2076e95B253F}{``bloXroute: Regulated Builder''}. This consistency demonstrated a critical finding: these block producers appear to follow rigorously base censorship but fail to scrutinize transactions involving addresses that previously received sanctioned funds (derived censorship). This highlights a significant gap in their current enforcement mechanisms. 

The remaining 8 block producers only partially complied or did not enforce \ac{OFAC} sanctions at all, such as \href{https://etherscan.io/address/0x4838B106FCe9647Bdf1E7877BF73cE8B0BAD5f97}{``Titan Builder''} which contributed the largest proportion (32.24\%) of blocks in which exist transactions that directly interacted with \ac{TC}. Meanwhile, \href{https://etherscan.io/address/0x95222290DD7278Aa3Ddd389Cc1E1d165CC4BAfe5}{``beaverbuild''} led in the number of blocks meeting certain impurity conditions, aligning proportionally with its overall block production. 

Furthermore, block producers derive significant revenue beyond transaction fees and block rewards from \ac{MEV}. In this process, \ac{MEV} bots pay block producers for favorable transaction placement, enabling profitable strategies such as arbitrage~\cite{DBLP:conf/sp/QinZG22}. Our investigation revealed that some of these \ac{MEV} operations were funded by \ac{TC}, as evidenced by bots like \href{https://etherscan.io/address/0x06e839b63191a6ee71b8f0907b52b89b082ac9fc}{\texttt{0x06e8...c9fc}} and \href{https://etherscan.io/address/0x000000009B801E488378Ae17EEC3c7c2BA73266C}{\texttt{0x0000...266C}}. This further raises critical compliance concerns, as block producers may unintentionally facilitate the movement of sanctioned assets, exposing themselves to regulatory risks.

Finally, gaps in consensus layer censorship lead the result that the assets derived from sanctioned addresses had been integrated into the entire Ethereum ecosystem. We conducted a comprehensive analysis on approximately 67.94 million addresses holding funds that could have originated from \ac{TC} at $B_\textit{END}$. Table~\ref{tab:impurity_distribution} summarizes the distribution of addresses and impurity amounts according to the impurity scores. For our initial analysis, we set the impurity threshold $\Phi = 5\%$. However, it is important to note that this threshold is not a definitive recommendation but serves primarily as an experimental baseline for Section~\ref{sec:insights}, as evaluating its efficacy requires ground truth data which is not universally available. The selection of an appropriate threshold involves inherent trade-offs that merit careful consideration. In practice, regulators and compliance teams should calibrate this parameter according to their operational constraints, risk tolerance, and available resources (potentially in conjunction with complementary metrics to enhance accuracy).

\begin{table}[tb]
\centering
\caption{Distribution of Addresses with their Impurity Amounts and Total Balance Amount by Impurity Scores $\varphi$}
\vspace{-8pt}
\resizebox{\linewidth}{!}{%
\begin{tabular}{lrrr}
\toprule
\textbf{Score} $\varphi$ & $\#$ \textbf{Address} & \textbf{Impurity Amount} $\mathcal{I}$ & \textbf{Total Balance Amount} $\mathcal{B}$ \\ 
\midrule
$[\, 0\%,  \   5\% \,)$  & 67,321,336 (99.08\%)  &  877,243.20 ETH (76.05\%)  & 127,696,284.34 ETH (99.62\%)  \\[4pt]

$[\, 5\%,  \  50\% \,)$  &    560,858 (00.83\%)  &    19,363.70 ETH (01.68\%)  &      224,987.50 ETH (00.18\%)  \\
\midrule
$[\, 50\%, \  95\% \,)$  &     16,665 (00.02\%)  &    4,155.41 ETH (00.36\%)  &       5,832.83 ETH (00.00\%)  \\[4pt]
$[\, 95\%, \ 100\% \,]$  &     45,258 (00.07\%)  &  252,763.76 ETH (21.91\%)  &     252,812.50 ETH (00.20\%)   \\
\midrule
\textbf{Total}                                  & 67,944,117 \ \, (100\%)    & 1,153,526.07 ETH \ \, (100\%)    & 128,179,917.19 ETH \ \, (100\%)    \\ 
\bottomrule
\end{tabular}
}
\label{tab:impurity_distribution}
\end{table}

\subsection{Gaps in Application Layer Censorship}
\label{subsec:application_gap}

While the consensus layer determines whether a transaction is included in the blockchain, most sanction controls sit at the application layer and many of them are bypassable.

\begin{subsubsection}{Off-Chain and On-Chain Censorship} 
We separate application layer censorship into off-chain and on-chain forms because the enforcement points and failure modes differ in practice. 

Off-chain censorship includes front-ends that hide functions or decline to provide services~\cite{khatri2022uniswap,khatri2022aave}, \ac{RPC} gateways that reject raw transaction submissions~\cite{avan-nomayo2022tornadocash}, and custodial venues that decline deposits or withdrawals under internal compliance rules. However, these decisions do not alter protocol semantics and are therefore non-binding with the blockchain state. A blocked user can switch to another front-end, send the transaction through a permissive \ac{RPC} endpoint, or interact with the contract directly using a \ac{CLT}.

On-chain censorship refers to logic that is written into smart contracts and evaluated at call time. We use the same definition in Wahrst{\"a}tter et al.'s work ~\cite{DBLP:conf/www/WahrstatterEYZQ24} as follows:

\begin{definition}[On-Chain Application Layer Censorship]
    A transaction $tx$ is censored by a smart contract, if $tx \in B_i$ , where $B_i \in \mathcal{L}$ is blocked by the state $s_i$ , s.t.~further state transitions $\mathcal{T}(s_i) \rightarrow s_{i+1}$ are blocked by the respective contract.
\end{definition}

\noindent The contract itself decides to refuse the state transition when a policy predicate is met, for example, the address is included in its denylist. It provides a clear causal link between the rule and the observed execution outcome such that users cannot bypass it using similar methods against off-chain censorship. However, because of the blockchain transparency, alongside with longer censorship paths (e.g., multisig confirmations and additional on-chain transaction proposal), an informed counterparty can move sanctioned assets before a blacklist update takes effect.

In the remainder of this subsection, our analysis concentrates on on-chain censorship implemented by the blacklist mechanism \textit{(i)}~Chainalysis Sanctions Oracle and \textit{(ii)} major stablecoin contracts (notably USDT and USDC). The evaluation of off-chain censorship effectiveness can be indirectly derived from downstream service utilization in Section~\ref{subsec:usage_of_service_providers}.
\end{subsubsection}

\begin{subsubsection}{Delays in Performing On-Chain Censorship} We tracked two quantities to measure delays. First, the policy-to-enforcement interval: the elapsed window between the official \ac{OFAC} publication time of a designation and the first on-chain transaction that adds the affected address to a blacklist. Second, the intra-block ordering position of that enforcement transaction: its index within the containing block, normalized by block length, which indicates how much ordering slack existed for adversaries to move first.

\begin{figure}[tb!]
\centering
\includegraphics[width=\columnwidth]{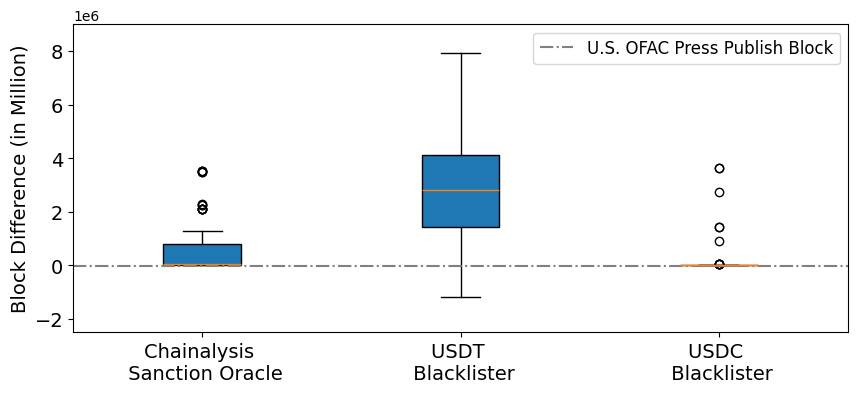}
\caption{The distribution of block height differences in millions between the first on-chain enforcement transaction and the reference block aligned to the \ac{OFAC} press-release block number (dashed zero line). Among three enforcement entities, USDC blacklister exhibits the shortest and least variable policy-to-enforcement delay.}
\Description{The distribution of block height differences in millions between the first on-chain enforcement transaction and the reference block aligned to the OFAC press-release block number (dashed zero line). Among three enforcement entities, USDC blacklister exhibits the shortest and least variable policy-to-enforcement delay.}
\label{fig:delay}
\end{figure}

For the policy-to-enforcement delay, we selected the block number closest to midnight of the OFAC publication date in Eastern Time Zone as $B_{\textit{OFAC}}$ and the block number of the first transaction that updates the blacklist as $B_{\textit{enforce}}$. We report $\Delta_{\textit{policy}} = B_{\textit{enforce}} - B_{\textit{OFAC}}$ in block numbers in Figure~\ref{fig:delay}, which represents the operator's response time plus any procedural latency. The result shows that USDC blacklister updates land essentially at the \ac{OFAC} publication time, with the Chainalysis Sanctions Oracle showing small, positive lags. By contrast, USDT blacklister exhibits much larger and more dispersed delays (but sometimes even appearing before), indicating slower and less consistent on-chain censorship.

For the intra-block position, let the enforcement transaction appear at index $i$ (zero-based) in a block with $N$ transactions. We computed a normalized rank $r = \frac{i}{N} \in [\, 0, 1 \,]$, where an $r$ near $1$ means the update landed late in the block; users with higher-priority fees could have been executed earlier in the same block. From our analysis, USDC blacklist transactions tend to be mined early in the block (avg. index \textbf{29.83\%}), reducing intra-block exposure. USDT transactions are appended mid-block (\textbf{47.01\%}), and Chainalysis oracle additions occur much later (\textbf{63.62\%}), leaving more room for same-block preemption by outgoing transfers that pay higher fees.
\end{subsubsection}

\subsection{Differences in User-Level Reaction}
\label{subsec:user_reaction}

Sanctions not only have impacts on enforcement, they also change how users route and time their transactions. Therefore, we further examine the on-chain behavioral adjustments. 

\begin{subsubsection}{Reduction in During-Sanction Activity}

\begin{table}[tb]
\caption{Impact of the Sanction on \ac{TC}: while withdrawal volume also declined significantly~(-\TCWithdrawVolumeDecreasePercentage), they remained slightly higher than deposit volumes~(-\TCDepositVolumeDecreasePercentage), suggesting users prioritized exiting over continued use.}
\vspace{-8pt}
\label{tab:sanction_impact}
\centering
\resizebox{\linewidth}{!}{%
\begin{tabular}{lcrrr}
\toprule
\textbf{Metric} & \textbf{Action} & \textbf{Pre-$B_\textit{OFAC}$} & \textbf{Post-$B_\textit{OFAC}$} & \textbf{Change} \\
\midrule
\multirow{2}{*}{\# Transaction}  
    & Deposit  & \TCDepositPreSanTxCount    & \TCDepositPostSanTxCount    & -\TCDepositTxCountDecreasePercentage \\
    & Withdraw & \TCWithdrawPreSanTxCount   & \TCWithdrawPostSanTxCount   & -\TCWithdrawTxCountDecreasePercentage \\ 
\midrule
\multirow{2}{*}{\# Address}  
    & Deposit  & \TCDepositPreSanAddrCount  & \TCDepositPostSanAddrCount  & -\TCDepositAddrCountDecreasePercentage \\
    & Withdraw & \TCWithdrawPreSanAddrCount & \TCWithdrawPostSanAddrCount & -\TCWithdrawAddrCountDecreasePercentage \\
\midrule
\multirow{2}{*}{Volume (in ETH)}       
    & Deposit  & \TCDepositPreSanVolume     & \TCDepositPostSanVolume     & -\TCDepositVolumeDecreasePercentage \\
    & Withdraw & \TCWithdrawPreSanVolume    & \TCWithdrawPostSanVolume    & -\TCWithdrawVolumeDecreasePercentage \\
\bottomrule
\end{tabular}
}
\end{table}

We extended the previous study of Wang et al.~\cite{DBLP:conf/www/WangCQZGBLG23} by examining a comprehensive dataset spanning \TCPreSanctionPeriod before and \TCSanctionPeriod after the \ac{OFAC} sanctions on \ac{TC}. Table~\ref{tab:sanction_impact} shows a substantial reduction in the number of transactions, unique addresses, and total volumes for both \ac{TC} deposits and withdrawals. Specifically, post-sanction deposit volume fell to only \textbf{28.97\%} of pre-sanction levels when comparing near equivalently long time periods. Figures~\ref{fig:temporal_analysis} displays the daily statistics for deposits and withdrawals. Following the announcement of the sanctions, we observed a significant increase in withdrawal transactions (labeled as Peak~\circledP). This peak represents the highest daily withdrawal transaction count after \ac{OFAC} sanctions and also indicates a mass exodus from \ac{TC}, suggesting that users were prompted to withdraw their assets due to heightened uncertainty.

\begin{figure}[tb!]
    \centering
    \includegraphics[width=\columnwidth]{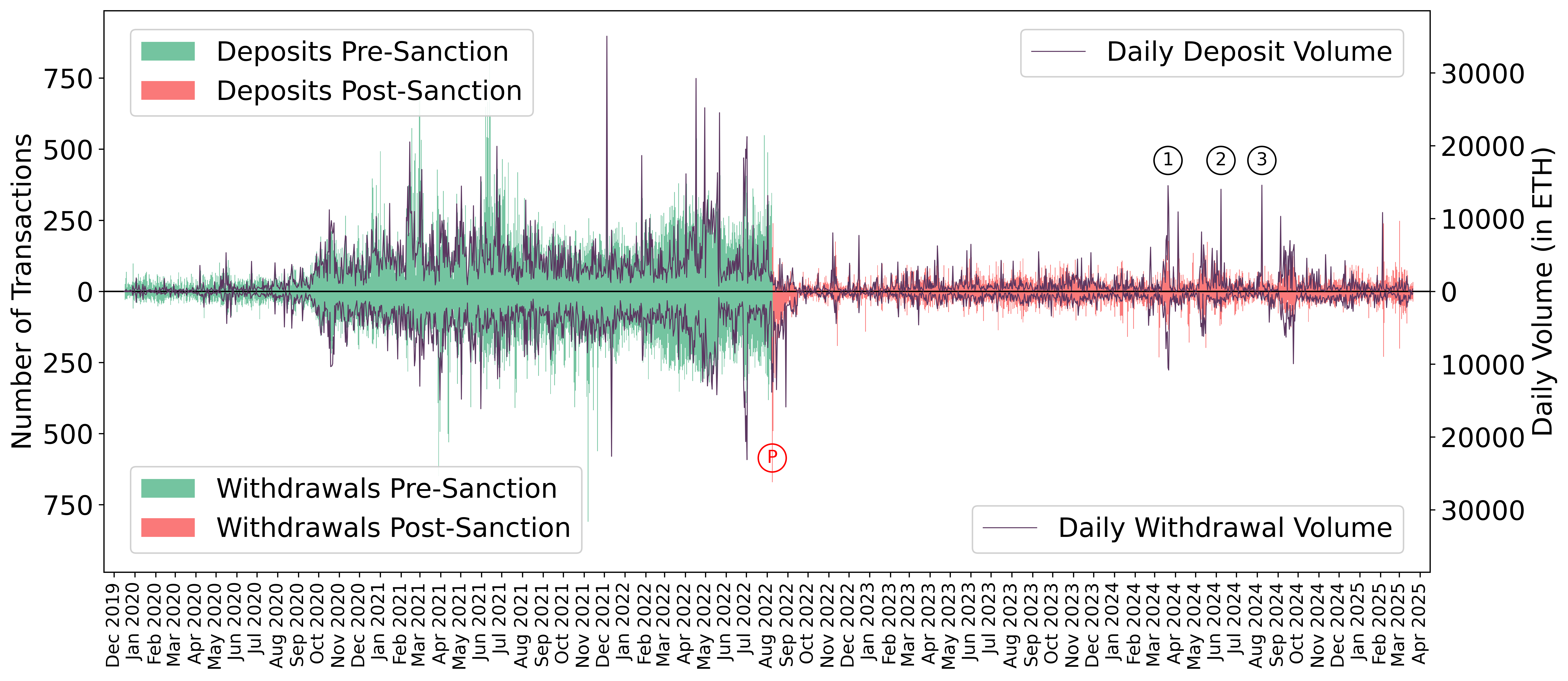}
    \caption{Daily transactions and volumes into the \ac{TC} ETH pools from \TCEstablishDate to \TCRemoveSanctionDate. The x-axis denotes the timeline, while the y-axis captures transaction count and volume of deposits (up) and withdrawals (down). The three circled values mark the top-3 during-sanction daily deposit volumes, each linked to a known security incident. Moreover, a significant spike in withdrawals is observed on the sanction date, marked by the circled letter ``P''.}
    \Description{Daily transactions and volumes into the TC ETH pools from December 16, 2019 to March 21, 2025. The x-axis denotes the timeline, while the y-axis captures transaction count and volume of deposits (up) and withdrawals (down). The three circled values mark the top-3 post-sanction daily deposit volumes, each linked to a known security incident. Moreover, a significant spike in withdrawals is observed on the sanction date, marked by the circled letter ``P''.}
    \label{fig:temporal_analysis}
\end{figure}
\end{subsubsection}

\begin{subsubsection}{Limited Effect on Ethereum Attackers}
\label{sec:limited_restriction_to_security_incidents}
We took a snapshot of the BlockSec database~\footnote{\url{https://docs.blocksec.com/phalcon/security-incident-list}} as of \EndStudyTime. This dataset catalogs significant \ac{DeFi} exploits with losses exceeding 100,000 USD and provides one of the most comprehensive public records of major attacks. Of the \textbf{80} total incidents on Ethereum in the dataset, funds were returned to victims in only 9 cases.  Among the remaining \textbf{71} malicious attacks, attackers in 11 cases retained stolen assets by converting them to non-freezable assets (e.g., ETH or DAI). Notably, among the \textbf{60} incidents with active money laundering activity, attackers used \ac{TC} as their primary laundering method in \textbf{47 cases (78.33\%)}. Moreover, we observed that in \textbf{35 (49.30\%)} out of these \textbf{71} malicious incidents, attackers sourced their funding from \ac{TC} to facilitate the exploit transactions. Importantly, \textbf{29 (82.86\%)} of these \textbf{35} cases subsequently returned stolen assets back into \ac{TC}, further obfuscating their on-chain footprints.

Moreover, we identified three significant deposit peaks after the sanction date, as illustrated in Figure~\ref{fig:temporal_analysis}. The first peak, denoted as Peak~\circledOne, occurred on March 21, 2024, predominantly driven by the \href{https://etherscan.io/address/0x7bEfDBB89C21863E910310A36Da5058704552935}{Heco Bridge Exploiter}~\cite{rekt_heco_bridge}, who contributed 11,200.00 ETH, representing \textbf{77.11\%} of the total deposits (14,522.00 ETH) on that day. The second peak, Peak~\circledTwo, took place on June 8, 2024, with deposits primarily originating from the \href{https://etherscan.io/address/0x84Bfa13778954ED29Db486fD2ef9acC7bA24773F}{Orbit Bridge Exploiter}~\cite{rekt_orbit_bridge}. This attacker deposited 12,930.00 ETH, accounting for \textbf{92.18\%} of the day's total deposit volume (14,027.00 ETH). The third and most pronounced peak, Peak~\circledThree, was recorded on August 8, 2024, with the \href{https://etherscan.io/address/0x663aD6B822537d1E8108436A80E4C1CCA4CCF448}{Nomad Bridge Exploiter}~\cite{rekt_nomad_bridge} depositing 14,509.30 ETH, constituting nearly the entirety (\textbf{99.50\%}) of the day's total 14,581.70 ETH.
\end{subsubsection}

\begin{subsubsection}{Dusting Attacks}
\label{subsubsec:dusting}
Following the announcement of \ac{OFAC} sanctions on \ac{TC}, a significant dusting attack emerged: small amounts of ETH were deliberately withdrawn from \ac{TC} to a wide range of addresses. Our analysis focused on transactions from the smallest \ac{TC} 0.1 ETH mixing pool within \DustAttackBlockCount blocks (approximately \DustAttackTime) after $B_\textit{OFAC}$. The dusting targets included prominent entities like \href{https://etherscan.io/tx/0xe8194847f1ab1ec97dde9a3b54cfa0d3cb63873a10e3af2dbe807251aaa05861}{Ethereum Foundation}, \acp{CEX} (e.g., \href{https://etherscan.io/tx/0x6ad84a551c838b8c6e12e923d18cdce7178b273ddecf7875d1dc4e285ce89c90}{Kraken}, \href{https://etherscan.io/tx/0x9c308e206c457ac42f09e5b1ad7f701db4a05917c85ed2790f79975a24b64f4f}{Binance}, \href{https://etherscan.io/tx/0xd4add05f3b9d2860df0096b113211a04938f0a478667ad28a5bc1b71360549d5}{Bitfinex}, \href{https://etherscan.io/tx/0x1574449cc28c699e921ee82dd39dea4bd32e3c9e6670813155fc1acf7a3fef14}{Gemini}, \href{https://etherscan.io/tx/0x72b86ebdf52e9bb9c13d485cd78ee72cd5cbbe4ac68a770fe9be544d7fba825c}{OKX}, etc.), and users interacting with \ac{DeFi} platforms. These dusting attacks effectively flagged numerous unsuspecting recipients as potentially sanctioned or involved in malicious activities.

A direct consequence of these dusting behaviors was the Denial-of-Service attack on services (e.g., Aave Front-End~\cite{khatri2022aave}) that are relying on binary classification mechanisms. We observed that dusters used merely \DustAttackUsedPreciseETH to taint at least \DustAttackTaintedETH (over \DustAttackAmplify) belonging to the affected recipients. This significantly increased the risk of asset freezing and service denial by compliant platforms. The dusting phenomenon underscores the need for continuous and quantitative risk metrics, such as our proposed impurity scores $\varphi$, that can capture varying levels of exposure to sanctioned funds. Such metrics enable regulators and service providers to implement proportional compliance responses based on precise risk assessments, rather than na{\"i}ve binary classifications. We also provide detailed arguments in Appendix~\ref{appendix:binary_impossibility}.
\end{subsubsection}


\section{Downstream Fund Flow Analysis}
\label{sec:insights}

In this section, we present detailed insights into how funds flow after withdrawals from sanctioned addresses. Specifically, we focus on two aspects: first, behavioral patterns prior to depositing into virtual asset services; and second, their subsequent interactions with these service providers. Through this analysis, we aim to enhance the understanding of user strategies and support efforts in tracing sanction-related activities across blockchain ecosystems.

\subsection{Behavioral Patterns}
\label{subsec:behavioral_pattern}

Between initial withdrawals and subsequent deposits into Ethereum services, we identify and analyze three primary behavioral patterns.

\begin{subsubsection}{Time of Initial Asset Transfers}
Regarding the time of asset transfers, our observations in Figure~\ref{fig:timeliness_of_asset_transfers} show significant variations in the duration between the first withdrawal of funds from \ac{TC} and subsequent movements of assets.

\begin{figure}[tb]
    \centering
    \includegraphics[width=\columnwidth]{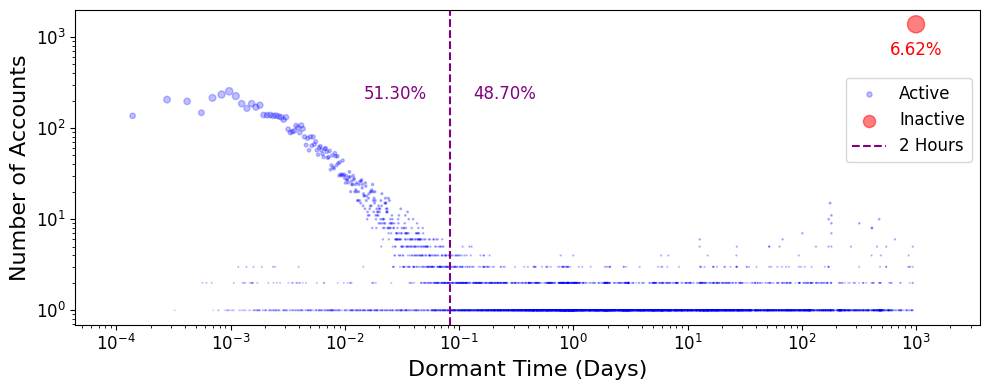}
    \caption{Dormant Time Distribution of \ac{TC} Withdrawers. $51.30\%$ of withdrawers transfer assets within $2$ hours after their first \ac{TC} withdrawal, while $48.70\%$ express a long-tail distribution pattern that ranges from several hours to hundreds of days. The red circle ($6.62\%$ of accounts) represents accounts that remain completely inactive.}
    \Description{Dormant Time Distribution of TC Withdrawers. 51.30\% of withdrawers transfer assets within 2 hours after their first \TC withdrawal, while 48.70\% express a long-tail distribution pattern that ranges from several hours to hundreds of days. The red circle (6.62\% of accounts) represents accounts that remain completely inactive.}
    \label{fig:timeliness_of_asset_transfers}
\end{figure}

The majority of withdrawers ($51.30\%$) transfer their assets within a $2$-hour window following the initial withdrawal, demonstrating a preference for immediate asset relocation. The remaining $48.70\%$ of withdrawers present extended dormant periods ranging from several hours to hundreds of days. This extended dormancy creates a long-tail distribution pattern visible in the scattered data points across the right side of the vertical threshold. The distribution of these delayed transfers appears non-uniform, with varying concentrations at different time intervals. The substantial cluster of inactive ones ($6.62\%$ of all \ac{TC} withdrawers) withdraw assets from TC and engage in no further transaction activity. 
\end{subsubsection}

\begin{subsubsection}{Split-and-Merge Pattern}
We applied our methods to both \ac{TC} withdrawers and the Bybit exploiter (the largest known security incident~\cite{elliptic_bybit_hack} in history until March 2025). Our comparative analysis, as shown in Figure~\ref{fig:motif_heatmap}, reveals the behavioral similarities between these actors based on the frequency of connected three-node motifs~\cite{doi:10.1126/science.aad9029}. The \emph{split-and-merge} pattern emerges as a distinctive signature, characterized by entities initially splitting funds across multiple addresses (Motif~021D), transferring them through intermediate addresses (Motif~021C), and finally merging assets (Motif~021U). This technique is specially designed to circumvent transaction monitoring systems and has been documented in both traditional and crypto-specific money laundering studies. This behavioral similarity, combined with widespread usage of \ac{TC} following security incidents (cf.~Section~\ref{sec:limited_restriction_to_security_incidents}), further raises concerns. 

\begin{figure}[tb]
    \centering
    \includegraphics[width=\columnwidth]{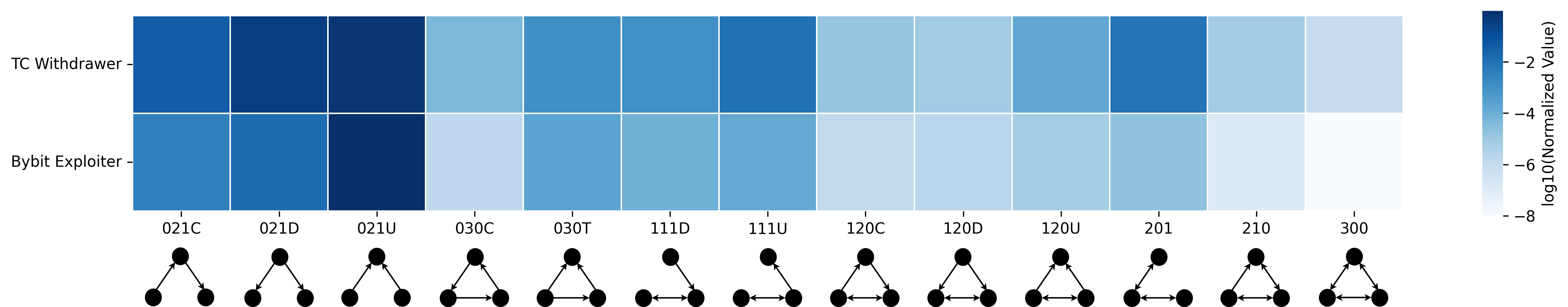}
    \caption{Heatmap comparing transaction patterns between \ac{TC} withdrawers and the Bybit exploiter based on connected three-node motifs. Dark blue cells of both distributions in 012C, 021D, and 021U show high-intensity activities of asset transferring, splitting, and merging, respectively.}
    \Description{Heatmap comparing transaction patterns between TC withdrawers and the Bybit exploiter based on connected three-node motifs. Dark blue cells of both distributions in 012C, 021D, and 021U show high-intensity activities of asset transferring, splitting, and merging, respectively.}
    \label{fig:motif_heatmap}
\end{figure}
\end{subsubsection}

\begin{subsubsection}{Test-Depositing Sequences}
Another behavior involves conducting small exploratory deposits before committing to substantial transfers. Table~\ref{tab:test_depositing} in Appendix~\ref{appendix:supplement} illustrates this behavior: a user first deposited around 0.1 ETH to KuCoin, then consistently transferred larger amounts (10--12 ETH) in all subsequent transactions.

This sequential approach serves the risk-mitigation purpose, where users perform an end-to-end operational check to ensure their funds can move without being blocked by any undisclosed sanctions control like the off-chain application layer censorship. Our analysis of five services implementing regulatory requirements reveals a precise proportion: 38.55\% of Binance, 75.31\% of Railgun, 49.02\% of OKX, 47.73\% of Gate.io, and 46.59\% of KuCoin users, who execute multiple deposits, demonstrate this test-then-deposit sequence. In contrast, platforms without sanctions enforcement exhibit different patterns. Users interacting with these services typically skip the testing phase, preferring to transfer substantial amounts in single transactions through primarily one-time intermediate addresses. This behavioral difference creates a detectable marker that compliance systems can leverage.
\end{subsubsection}

\subsection{Service Utilization}
\label{subsec:usage_of_service_providers}

Following our examination of behavioral patterns, we investigated the diverse service providers that \ac{TC} withdrawers commonly use.

\begin{subsubsection}{Cross-Chain Bridges}

\begin{figure}[tb!]
    \centering
    \includegraphics[angle=270, width=\columnwidth]{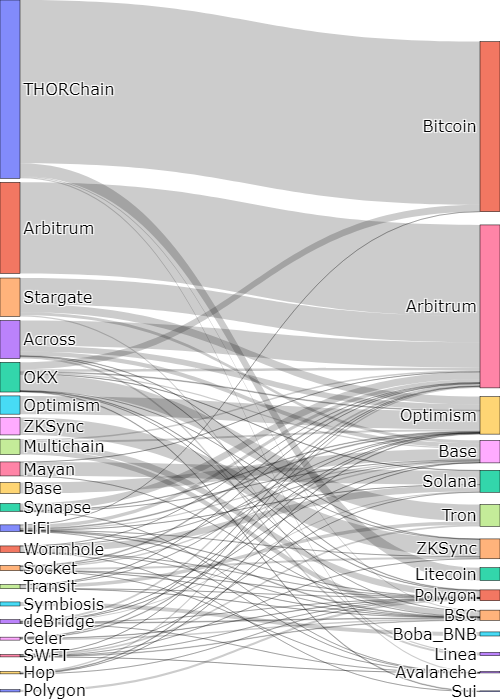}
    \caption{Flow of Funds via Cross-Chain Bridges. We implement custom parsers for 20 bridges and 2 proxy aggregators to get the corresponding target chain from the transaction metadata. Among over 15 destinations, Bitcoin ($33.31\%$) and EVM-compatible blockchains ($53.02\%$) are primary options.}
    \Description{Flow of Funds via Cross-Chain Bridges. We implement custom parsers for 20 bridges and 2 proxy aggregators to get the corresponding target chain from the transaction metadata. Among over 15 destinations, Bitcoin (33.31\%) and EVM-compatible blockchains (53.02\%) are primary options.}
    \label{fig:bridge}
\end{figure}

Cross-chain bridges have emerged as pivotal components within the cryptocurrency ecosystem, enabling users to seamlessly transfer assets and messages across disparate blockchain networks. Although these mechanisms facilitate interoperability and expand use cases, they also pose significant challenges to on-chain forensics and censorship efforts, as shown in Figure~\ref{fig:bridge}.

\textbf{Plaintext Metadata.} Most cross-chain protocols disclose core transaction parameters in plaintext, including the destination chain identifier and the recipient address on the destination chain. In principle, these metadata enable analyzers to link an outgoing transaction on one chain to its corresponding incoming transaction on another. However, a lack of standardization across bridging protocols, especially when destination chains are non-\ac{EVM}-compatible, hampers this process. Even identifiers for the same blockchain can vary between services built on top of both \ac{EVM}-compatible underlying infrastructures. For instance, \href{https://etherscan.io/address/0x77b2043768d28E9C9aB44E1aBfC95944bcE57931}{``Stargate: Pool Native''} uses 0x75e8 (30184) to represent the Base blockchain, while \href{https://etherscan.io/address/0x5c7BCd6E7De5423a257D81B442095A1a6ced35C5}{``Across Protocol: Ethereum Spoke Pool V2''} uses 0x2105 (8453). The former adopts the Endpoint ID defined by LayerZero\footnote{\url{https://docs.layerzero.network/v2/deployments/deployed-contracts}} and the latter uses the common Chain ID\footnote{\url{https://chainlist.org/}}. Such inconsistencies necessitate the development of custom parsers for each bridge implementation, significantly increasing the effort required to integrate, interpret, and verify metadata across platforms.

To address this challenge, we developed custom parsers for 20 widely used cross-chain bridges and 2 proxy aggregators (\href{https://etherscan.io/address/0xFc99f58A8974A4bc36e60E2d490Bb8D72899ee9f}{OKX: Web3 Proxy} and \href{https://etherscan.io/address/0x1231DEB6f5749EF6cE6943a275A1D3E7486F4EaE}{LI.FI: LiFi Diamond}) frequently involved in \ac{TC} withdrawals. Our analysis revealed that $\mathcal{I}_v = 350,918.15$ ETH ($\varphi \approx 97.37\%$) were bridged out of Ethereum via these platforms, contributing to \textbf{32.41\%} of total impurity volume. \ac{EVM}-compatible chains accounted for more than half (53.02\%) of the bridged impurity volume, including eight Layer-2 rollups and two alternative Layer-1 chains. Among heterogeneous targets, Bitcoin emerged as the dominant destination, capturing 33.31\% of bridged volume where 95.95\% of them were routed through THORChain. Solana and Litecoin accounted for an additional 4.30\% and 2.51\%, respectively. However, some bridges such as Chainflip ($\mathcal{I}_v = 4,601.81$ ETH, $\varphi = 97.80\%$) and Orbiter ($\mathcal{I}_v = 1952.27$ ETH, $\varphi = 93.23\%$) do not provide sufficient on-chain information, preventing identification of destination chains or recipient addresses via transaction analysis.

\textbf{Cross-Chain Motivations.} \emph{(i) Hard Fork Tolerance}: The DAO hard fork~\cite{DBLP:journals/jcit-igi/MeharSGGFSKL19} in 2016 highlighted Ethereum's capacity to reverse transactions through community-driven rollbacks. Because this outcome undermined finality for malicious actors, they would bridge funds to different chains, aiming to reduce the likelihood of reversal. \emph{(ii) Heightened Obfuscation:} Each newly designed blockchain or newly developed cross-chain bridge introduces additional complexity into the transaction tracing. This challenge is amplified when intermediary chains provide limited transparency or lack robust blockchain analytics tooling. Moreover, certain blockchains support built-in or composable privacy-enhancing mechanisms (e.g., confidential transactions on Monero, CoinJoin on Bitcoin and \ac{TC} on Ethereum). Users can strategically route funds through chains with advanced anonymity features to further obscure the provenance of assets. When privacy tools are chained together, e.g., bridging from Ethereum to Bitcoin for CoinJoin, then returning to Ethereum for \ac{TC}, the difficulty of end-to-end tracing increases exponentially. The limited regulatory or compliance enforcement on newly developed chains (e.g., Boba and Sui) also incentivizes cross-chain transfers for evading detection. \emph{(iii) Lower Fees and Faster Confirmations:} Layer-2 rollups and some Layer-1 blockchains offer significantly reduced transaction costs and faster block finality compared to Ethereum. While these properties benefit users seeking performance and affordability, they are equally attractive to malicious actors seeking to move funds cheaply and efficiently.
\end{subsubsection}

\begin{subsubsection}{Decentralized Exchanges}
\begin{figure}[tb!]
    \centering
    \includegraphics[angle=270, width=\columnwidth]{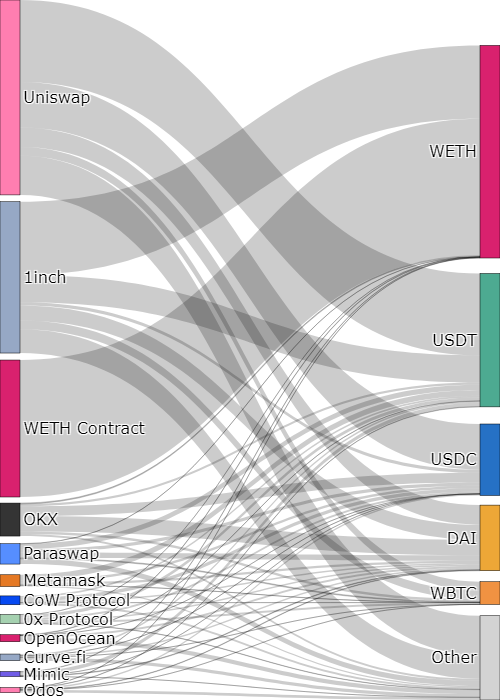}
    \caption{Flow of Funds via Decentralized Exchanges. Over $86.27\%$ of impurity volume are swapped into stablecoins and wrapped tokens. Notably, USDT and USDC have blacklist mechanisms to freeze assets held by sanctioned entities.}
    \Description{Flow of Funds via Decentralized Exchanges. Over $86.27\%$ of impurity volume are swapped into stablecoins and wrapped tokens. Notably, USDT and USDC have blacklist mechanisms to freeze assets held by sanctioned entities.}
    \label{fig:dex}
\end{figure}

\acp{DEX} are significant choices utilized by sanctioned addresses, accounting for \textbf{26.78\%} of all impurity volume $\mathcal{I}_v$ in our database. This preference stems from their permissionless architecture and censorship-resistant design.

\textbf{Dominant Service Providers.} We identified transactions flowing into more than 12 \ac{DEX} platforms as shown in the upper side of Figure~\ref{fig:dex}. Among these platforms, Uniswap~\cite{uniswap_protocol} (33.04\%) and 1inch~\cite{1inch_protocol} (25.64\%) dominated downstream \ac{DEX} activities. Although their official websites profess compliance with sanctions and implement front-end filters~\cite{khatri2022uniswap}, these measures can be circumvented through direct on-chain contract invocations via \ac{CLT}, and finally make off-chain application layer censorship ineffective. Consequently, large volumes of \ac{TC}-derived funds ($\mathcal{I}_v = 289,901.92$ ETH) still flowed into platforms. Specifically, we observed that 102,928.18 ETH (impurity score $\varphi \approx 93.04\%$) channeled through Uniswap and 77,481.93 ETH ($\varphi \approx 95.95\%$) processed via 1inch.

\textbf{Freezable vs. Non-Freezable Tokens.} After analyzing the transaction logs and account state differences, we found that \ac{TC} withdrawers frequently convert their assets into stablecoins (USDT-22.61\%, USDC-12.12\%, and DAI-11.12\%) or wrapped tokens (WETH-36.03\% and WBTC-3.85\%) for greater fungibility and cross-chain interoperability. These conversions from highly tainted ETH inputs result in comparably high impurity scores across all tokens. It is worth noting that USDT (Tether~\cite{tether}) and USDC (Circle~\cite{circle}) are centrally issued by U.S.-based companies. Their token contracts have explicit blacklist mechanisms, \verb|addBlackList (0x0ecb93c0)| and \verb|blacklist (0xf9f92be4)|, theoretically allowing privileged addresses to freeze assets held by sanctioned entities. However, our data showed no statistically significant difference in token selection patterns between potentially freezable tokens (USDT, USDC) and non-freezable alternatives (WETH, WBTC, DAI). This suggests that the theoretical risk of asset freezing does not substantially influence withdrawers' token selection strategies, highlighting a significant gap between technical capability and operational enforcement.
\end{subsubsection}

\subsubsection{Additional Analysis of Service Usages} Due to the space constraint, detailed analyses of \textit{(i)} centralized exchanges and \textit{(ii)} privacy-enhancing tools are moved to Appendix~\ref{appendix:additional}.


\section{Discussion}
\label{sec:discussion}

Sanction enforcement on blockchains is inherently a \textit{Cat-and-Mouse} game. As regulators develop detection and enforcement mechanisms, adversaries continuously adapt with sophisticated evasion strategies. For ethical considerations, no personally off-chain identifiable information is used and no actionable specific for circumvention is revealed. We carefully calibrated our disclosures which indicate that the current censorship framework has structural limitations. Here we also provide practical recommendations for compliance teams and regulated blockchain ecosystem developers: \textit{(i)}~cut policy-to-enforcement latency using fast off-chain enforcement first while updating a permanent on-chain state, \textit{(ii)}~provide a safe and user-initiated ``burn'' mechanism, so users accidentally exposed to tainted assets can destroy them, and \textit{(iii)}~standardize cross-chain bridge protocols so hops are auditable end-to-end.


\section{Related Work}
\label{sec:related}

Our work builds upon previous research on blockchain transaction analysis and privacy mechanisms. Prior works by Wu et al.~\cite{DBLP:journals/tifs/WuLFYCZS24} and Gomez et al.~\cite{DBLP:conf/ccs/GomezMC22} established graph-based approaches for tracking malicious funds from security incidents, while Lin et al.~\cite{DBLP:conf/www/0007WYFZY24} proposed density-based detection of money laundering patterns on Ethereum. Anonymity analyses in~\cite{DBLP:conf/www/WangCQZGBLG23, DBLP:conf/www/KovacsS24, DBLP:journals/tdsc/WahrstatterTS24, DBLP:journals/tifs/DuCSZH24, brownworth2024regulating} evaluated the capability of privacy-enhancing protocols using heuristic-based or machine-learning-based approaches. Yousaf et al.~\cite{DBLP:conf/uss/YousafKM19} first identified the complexity of cross-chain tracing between blockchains that have anonymous mechanism, while our research implements custom parsers for major bridge protocols.

Zola et al.~\cite{DBLP:conf/esorics/ZolaMO24} first showed the ineffectiveness of sanctions on Bitcoin, but limited their analysis at 2-step transaction graphs. M{\"o}ser et al.~\cite{DBLP:conf/fc/MoserBB14} also proposed a risk scoring mechanism on Bitcoin. However, the account model of Ethereum and the capability of smart contracts bring more mechanisms for both censorship and obfuscation. Wahrst{\"a}tter et al.~\cite{DBLP:conf/www/WahrstatterEYZQ24} documented censorship at the block production level, but lacked downstream money flow analysis.  Our work provides the first comprehensive empirical analysis of sanction effectiveness on Ethereum, introducing a quantitative framework that evaluates both direct compliance and subsequent circumvention patterns with high precision.


\section{Conclusion}
\label{sec:conclusion}

This paper presents the first quantitative evaluation of blockchain sanctions effectiveness.
We use \ac{TC} as a case study, analyzing 957 days of Ethereum data. Our findings reveal that while general \ac{TC} deposit volume decreased by 71.03\% during sanctions, malicious actors continued exploiting TC in 78.33\% of security incidents. This continued exploitation is facilitated by inconsistent consensus layer enforcement and sophisticated evasion techniques, including specific behavior patterns and strategic service selection. These insights provide concrete recommendations for regulatory authorities and compliance teams, helping them balance privacy concerns with financial oversight requirements using quantitative risk frameworks.


\newpage

\bibliographystyle{ACM-Reference-Format}
\balance
\bibliography{references}

\appendix

\section{Algorithm}
\label{appendix:algorithm}

In this section, we provide the algorithm for updating the impurity score of each affected address as sanctioned assets propagate.

\begin{algorithm}[h]
    \begin{small} 
    \caption{Impurity Updating Algorithm}
    \label{alg:impurity_updating_algorithm}
    \SetAlgoLined 

    \KwData{Blockchain states $\{(s_0, s_1, \dots, s_{n-1}) \, | \, s\in \mathcal{S} \}$, Valid Ethereum address set $\textit{ADDR}$, Sanctioned address set $\textit{ADDR}_\textit{sanction} \subset \textit{ADDR}$, First state after sanctions $s_{\textit{OFAC}}$}

    \KwResult{Historical records of for $\langle \mathcal{I}, \mathcal{B},\varphi \rangle$ all involved addresses at each state $s$ where $\mathcal{I}$ is the impurity amount, $\mathcal{B}$ is the balance amount, and $\varphi$ is the impurity score (cf. Section~\ref{subsec:metrics})}

    \textbf{Initialization:} \\
    $s_0 = s_\textit{OFAC}$\;
    \For{each address $addr \in \textit{ADDR}_\textit{sanction}$}{
        Set $\mathcal{I}(s_{\textit{OFAC}}, addr) = \mathcal{B} (s_{\textit{OFAC}}, addr)$\;
    }
    \For{each address $addr \notin \textit{ADDR}_\textit{sanction}$}{
        Set $\mathcal{I}(s_{\textit{OFAC}}, addr) = 0$\;
    }

    \textbf{State Transition and Attribute Update:} \\
    \For{$i \in [0, \, n-1)$}{
        Let balance state transition $\mathcal{T}_\mathcal{B}(s_i, op_\mathcal{B}) \rightarrow s_{i+1}$
        where affected addresses and amounts are extracted from the operation $op_\mathcal{B}$:\\
        $op_\mathcal{B} \rightarrow \{\langle addr_{\textit{from}}, a_{\textit{sent}}\rangle, \langle addr_{\textit{to}}, a_{\textit{received}}\rangle\}$\;

        \textbf{Update:} \\
        $\mathcal{I}(s_{i+1}, addr_{\textit{from}}) =
            \mathcal{I}(s_i, addr_{\textit{from}}) - \Bigl\lceil \frac{a_{\textit{sent}} \cdot \mathcal{I}(s_i, addr_{\textit{from}})}{\mathcal{B}(s_i, addr_{\textit{from}})}  \Bigr\rceil$\;

        \eIf{address $addr_{\textit{to}} \in \textit{ADDR}_\textit{Sanction}$}{
            $\mathcal{I}(s_{i+1}, addr_{\textit{to}}) = \mathcal{B}(s_{i+1}, addr_{\textit{to}})$\;
        }{  
            $\mathcal{I}(s_{i+1}, addr_{\textit{to}}) = 
                \mathcal{I}(s_i, addr_{\textit{to}}) + \Bigl\lceil \frac{a_{\textit{received}} \cdot \mathcal{I}(s_i, addr_{\textit{from}})}{\mathcal{B}(s_i, addr_{\textit{from}})}  \Bigr\rceil$\;
        }
        \For{each address $addr \in \big\{addr_{\textit{from}}, \, addr_{\textit{to}}\big\}$}{
            \eIf{$\mathcal{B}(s_{i+1}, addr) \neq 0$}{
                $\varphi(s_{i+1}, addr) = \frac{\mathcal{I}(s_{i+1}, addr)}{\mathcal{B}(s_{i+1}, addr)}$\;
            }{
                $\varphi(s_{i+1}, addr) = 0$\;
            }
            \textbf{Store}
            $\Bigl<  \mathcal{I}(s_{i+1}, addr), \, \mathcal{B}(s_{i+1}, addr), \, \varphi(s_{i+1}, addr) \Bigr>$
        }
    }
    \end{small}
\end{algorithm}

\section{Impossibility of Binary Classification Mechanism}
\label{appendix:binary_impossibility}
We provide an informal argument demonstrating why na{\"i}ve binary classification system inevitably fails when faced with dusting attacks, thereby necessitating the continuous, quantitative impurity metrics described in Section~\ref{subsec:metrics}. We also discuss the effectiveness of other classification mechanisms in this section.

Let $\textit{ADDR}$ be the set of all valid addresses on the blockchain, and define a general binary classification function:
\begin{equation}
 \mathcal{C}: \textit{ADDR} \rightarrow \{0,1\}   
\end{equation}
where $\mathcal{C}(addr) = 0$ means address $addr$ is ``clean'' and $\mathcal{C}(addr) = 1$ means $addr$ is ``tainted.'' Initially, sanctioned addresses are tainted: $\mathcal{C}(addr) = 1$ if $addr \in \textit{ADDR}_{\textit{sanction}}$, otherwise $\mathcal{C}(addr) = 0$.

For a binary classification system to be useful in blockchain analysis, it must satisfy two key properties:

\begin{enumerate}
    \item \textbf{Propagation Property}: There must exist some mechanism by which taint propagates from sanctioned addresses to their transaction partners. Without this property, the classification would fail to identify entities interacting with sanctioned addresses.
    \item \textbf{Decidability Property}: The classification must be computationally decidable in reasonable time, meaning there exists an efficient algorithm to determine whether an address is tainted.
\end{enumerate}

Any binary classification system satisfying these minimal properties must implement some form of propagation rule. While the specific rule may vary, it must establish conditions under which an address receiving funds from a tainted address also becomes tainted. We denote this general propagation condition:
\begin{equation}
    \mathcal{P}(s, addr_{\textit{from}}, addr_{\textit{to}}, v)
\end{equation}
which evaluates to true when a transaction from $addr_{\textit{from}}$ to $addr_{\textit{to}}$ with a value $v$ at the blockchain state $s$ causes taint to propagate.

A dusting attack occurs when an attacker controlling a tainted address $addr_s$ (where $C(addr_s) = 1$) creates transactions
\begin{equation}
    tx_i: addr_s \rightarrow addr_i \implies \mathcal{P}(s, addr_{\textit{s}}, addr_{\textit{i}}, v_i) = \textit{True}
\end{equation} with values $v_i$ deliberately chosen to ensure taint is propagated for many strategically selected target addresses $addr_i$. Let $G(s) \subseteq \textit{ADDR}$ be the set of all tainted addresses at blockchain state $s$, with $G(s_\textit{OFAC}) = \textit{ADDR}_\textit{sanctioned}$ initially. After each dusting attack, $G(s')$ expands to include the targeted addresses.

The fundamental limitation of na{\"i}ve binary classification system is that it cannot distinguish between significant and trivial taint exposure: an address is either tainted or not. This creates a vulnerability that can be exploited through strategic dusting: a small-value dusting transaction can instantly taint high-value addresses, creating an amplification effect. From our empirical data in Section~\ref{subsubsec:dusting}, just \DustAttackUsedPreciseETH in dusting transactions tainted over \DustAttackTaintedPreciseETH, an amplification factor exceeding \DustAttackAmplify.

We denote the total value (balance amount) held by tainted addresses flagged by na{\"i}ve binary classification mechanism as 
\begin{equation}
    V_\text{tainted}(s) = \sum_{addr \in G(s)} \mathcal{B}(s, addr)
\end{equation} and denote the total value held by entire addresses as 
\begin{equation}
    V_\text{entire}(s) = \sum_{addr \in \textit{ADDR}} \mathcal{B}(s, addr)
\end{equation} Under propagation rules satisfying the minimal properties, strategic dusting attacks to na{\"i}ve binary classification system can drive 
\begin{equation}
\begin{aligned}
    \lim_{s \to \infty} \frac{|\,G(s)\,|}{|\, \textit{ADDR} \,|} &\to 1 \\
    \lim_{s \to \infty} V_\text{tainted}(s) &\to V_\text{entire}(s) 
\end{aligned}
\end{equation}
When most addresses become tainted, the system cannot distinguish genuine risk from dust-induced taint, rendering the classification meaningless. Services implementing na{\"i}ve binary classification face an inevitable denial-of-service vulnerability: either block all tainted addresses (causing service failure as $|G(s)|$ grows) or ignore the classification (rendering it useless). This fundamental limitation was observed with services like Aave Front-End~\cite{khatri2022aave}.

The failure of na{\"i}ve binary classification system directly necessitates a continuous metric such as our impurity score $\varphi(s, addr) \in [0,1]$. Under dusting attacks, na{\"i}ve binary classification marks all recipients as completely tainted, while with our continuous metric, $\varphi(s, addr_i) = \frac{\mathcal{I}(s, addr_i)}{\mathcal{B}(s, addr_i)} \approx 0$ for addresses with large balances receiving small dust amounts. This allows the system to distinguish between seriously tainted addresses and those with negligible exposure, preserving the utility of the classification system against dusting attacks. Our impurity updating algorithm in Appendix~\ref{appendix:algorithm} implements precisely this approach.

In the following of this section, we will discuss the effectiveness of other classification rules from aspect of attackers and evaders: 

\begin{enumerate}
\item \textbf{For Time-Based Rules} (e.g., taint expires after time $t$): \textit{(i)} Attackers can refresh the dusting attack periodically; \textit{(ii)} Evaders can bypass the detection via a waiting time longer than $t$.

\item \textbf{For Hop-Based Rules} (e.g., taint if receiving from any address that is within $h$ hops of tainted addresses): \textit{(i)} Attackers can directly issue transactions from tainted addresses to target addresses in $1$ hop; \textit{(ii)} Evaders cannot bypass unless they can move assets through addresses or services that reset or do not share the hop count, and eventually exceed the hop limit $h$.

\item \textbf{For Value-Threshold Rules} (e.g., taint if receiving more than $\theta$ value from sanctioned addresses): 

\begin{itemize}
    \item \emph{If the detection only check the value in target addresses:} \textit{(i)} Attackers can simply send $v > \theta$ in one transaction or analyze on-chain data to send more precision value; \textit{(ii)} Evaders cannot bypass unless they send the amount exceeding $\theta$ to a designated burnt address or divide their balance through services do not apply this detection mechanism.
    \item \emph{If the detection also check the value in transactions:} \textit{(i)} Attackers can divide $v > \theta$ into several transactions and send them sequently; \textit{(ii)} Evaders' operations do not change.
    \item \emph{If the detection even check the value in source addresses:} \textit{(i)} Attackers can prepare several addresses totally holding $v > \theta$; \textit{(ii)} Evaders' operations do not change.
\end{itemize}

\item \textbf{For Percentage-Threshold Rules} (e.g., taint if receiving more than $\theta$\% of balance from sanctioned addresses):

\begin{itemize}
    \item \emph{If the detection only check the percentage in target addresses:} \textit{(i)} Attackers can simply send $v > \theta\% \times \mathcal{B}(s,addr_i)$ in one transaction or analyze on-chain data to send more precision value; \textit{(ii)} Evaders can supply their account with enough ``clean'' amount to decrease the percentage $\varphi < \theta\%$. This can be done by direct transfer or using balance manipulation (e.g., addresses can drain and restore their balance using services do not apply this detection mechanism, for example, flash loans via \ac{DEX} and coin mixing via privacy tools).
    \item \emph{If the detection also check the percentage in transactions:} \textit{(i)} Attackers can prepare one or more accounts where each account is under the threshold $\varphi < \theta\%$ and collectively send $v > \theta\% \times \mathcal{B}(s,addr_i)$; \textit{(ii)} Evaders' operations do not change.
    \item \emph{If the detection even check the percentage in source addresses:} Both attackers' and evaders operations do not change.
\end{itemize}
\end{enumerate}

To conclude, na{\"i}ve binary rules are fundamentally untenable under dusting attacks. We advocate continuous metrics with calibrated thresholds and contextual checks (value, time, hops) to enable enforcement that remains robust to dust and evasion tactics.

\section{More Analyses of Service Usage}
\label{appendix:additional}

\subsection{Centralized Exchanges}

\acp{CEX} serve as significant on- and off-ramps which bridge the gap between digital and fiat currencies. Their ability to convert large sums of crypto into fiat and vice versa makes them pivotal in off-chain \ac{AML} and \ac{CTF} processes. Despite the potential exposure to increased surveillance risks, \acp{CEX} remain a frequent choice (\textbf{10.37\%}) among \ac{TC} withdrawers.

\textbf{\ac{KYC} vs. Non-\ac{KYC} Compliance.} A \ac{CEX} that follows \ac{KYC} compliance typically mandates identity verification procedures, such as submitting government-issued identification, to fulfill regulatory requirements. Users are generally required to pass these checks to deposit or withdraw fiat currencies, and often when transacting significant volumes of assets. However, in practice, many exchanges, which are nominally \ac{KYC}-compliant, only require partial identity verification, usually triggered when users interact with fiat gateways. In these cases, users who fail to provide the necessary documentation may face service restrictions or be forced into refund mechanisms, particularly when their behavior raises compliance risks (e.g., as observed on platforms like ChangeNOW \footnote{\url{https://changenow.io/en/terms-of-use/changenow-terms}} and SimpleSwap\footnote{\url{https://simpleswap.io/terms-of-service}}). To better reflect real-world behavior, we extend the definition of non-\ac{KYC}-compliant \acp{CEX} to include those platforms that do not require personal information for account registration and for crypto-to-crypto transactions.  In such settings, linkabilities \textit{(i)} between deposit and withdrawal interactions with \ac{CEX} is broken when analyzing on-chain information, and \textit{(ii)} between CEX's internal accounts and off-chain user identifiers is also broken. As a result, once funds move through these non-\ac{KYC}-compliant \acp{CEX}, tracing them becomes highly improbable unless these exchanges record the transaction history and are willing to share them.

\begin{figure}[b]
\centering
\includegraphics[width=0.8\columnwidth]{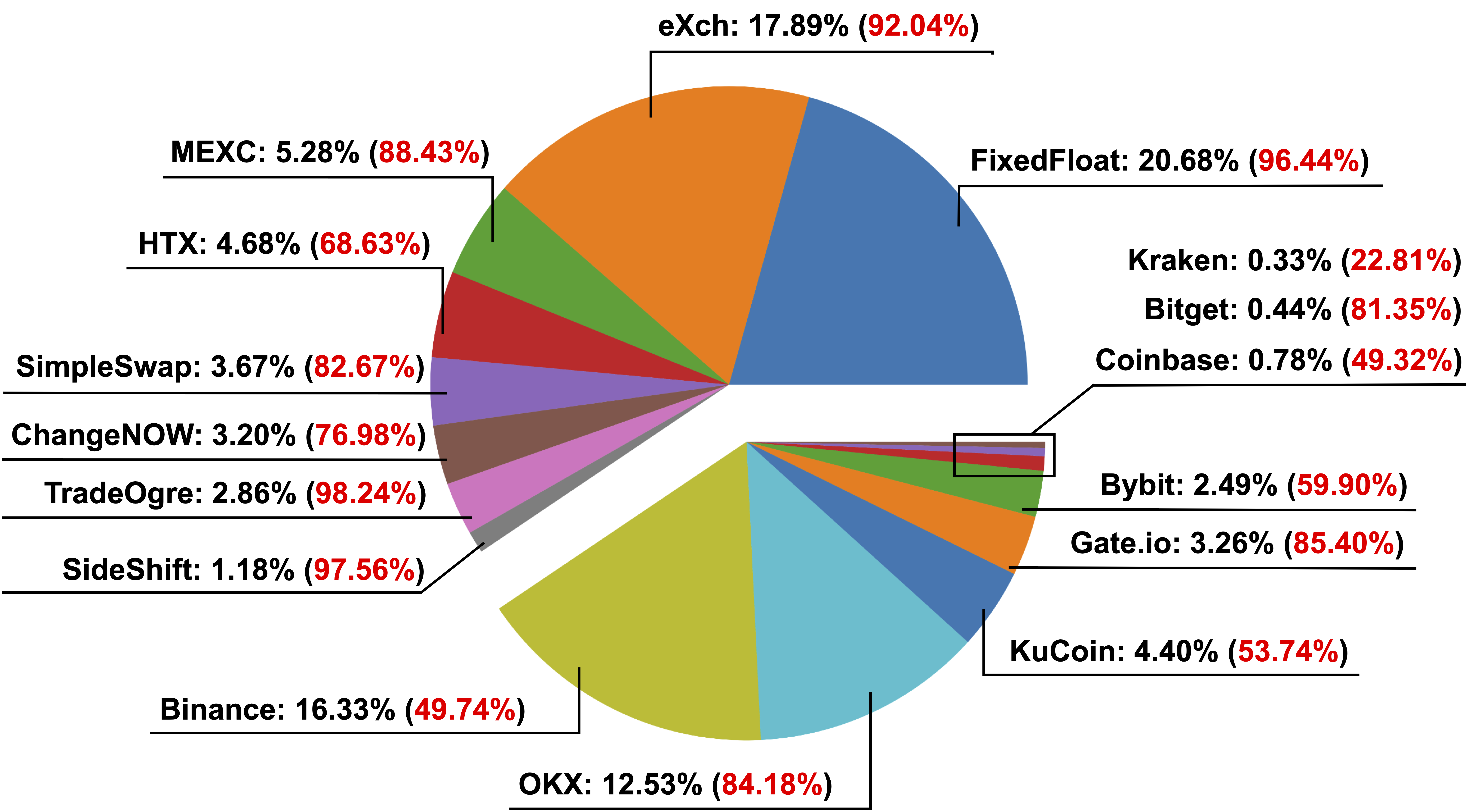}
\caption{The distribution of tainted funds deposited into selected \acp{CEX}, categorized by 8 Non-\ac{KYC} platforms (up) and 8 \ac{KYC}-compliant ones (down). The black percentages are proportions of impurity volumes into these platforms, while the red percentages reflect the corresponding impurity scores.}
\Description{The distribution of tainted funds deposited into selected CEXs, categorized by 8 Non-KYC platforms (up) and 8 KYC-compliant ones (down). The black percentages are proportions of impurity volumes into these platforms, while the red percentages reflect the corresponding impurity scores.}
\label{fig:cex_distribution}
\end{figure}

\textbf{Impurity Volume, Impurity Score and Active Spans.} We analyzed 16 \acp{CEX} in total, 8 \ac{KYC}-compliant and 8 non-\ac{KYC}-compliant, focusing on three key metrics: impurity volume $\mathcal{I}_v$, impurity score $\varphi$, and active spans (the number of days on which deposits with $\varphi > \Phi$ are detected). Figure~\ref{fig:cex_distribution} and Table \ref{tab:cex_volume_and_span} summarize our findings. Several non-\ac{KYC}-complaint \acp{CEX} (e.g., ChangeNOW, FixedFloat, and eXch, highlighted in Table~\ref{tab:cex_volume_and_span}) exhibit either a larger number of active days or a higher impurity volume compared to top-tier \ac{KYC}-compliant \acp{CEX} (e.g., Coinbase and KuCoin). We also observed higher impurity scores for deposits into non-\ac{KYC}-compliant \acp{CEX} relative to their \ac{KYC}-compliant counterparts. The lack of stringent identity checks reduces the risk of immediate account suspension or asset freezing, making these Non-\ac{KYC} platforms more attractive. 

Notably, U.S.-based platforms (i.e., Coinbase and Kraken) demonstrate relatively low impurity amounts and scores. This pattern is likely attributable to strict regulatory under the \ac{OFAC} sanctions targeting \ac{TC}, which encourages stricter transaction screening and discourages high-risk deposits. Conversely, due to the complexity of cross-jurisdictional enforcement, some \ac{KYC}-complaint \ac{CEX} based outside the U.S. (e.g., OKX and Bybit) may not follow the specific OFAC sanctions. Therefore, their impurity scores resembled those of non-\ac{KYC}-compliant platforms, even though their official websites claim compliance with \ac{AML} and \ac{CTF} policies.

\subsection{Privacy-Enhancing Tools}
Multiple privacy tools have emerged within the blockchain ecosystem,  each employing distinct techniques to break the linkability between transactions.  An advanced strategy observed in privacy-focused activities involves the sequential use of multiple privacy protocols, either at different stages of the transaction lifecycle or across different blockchains.

\textbf{Reuse of Tornado Cash.} \ac{TC} itself also serves as a destination for assets previously withdrawn from its own pools. Our analysis identifies two prevalent patterns of reuse: \emph{(i) Self-Loop}, where users withdraw funds from \ac{TC} and redeposit nearly identical amounts back into \ac{TC}, sometimes routing these assets through one or more intermediate addresses; and \emph{(ii) Mix-up}, where withdrawn assets from \ac{TC} are combined with external funds, thereby reducing the overall impurity score. In total, we detect 54,249.31 ETH involved in such reuse cycles, resulting in an impurity score $\varphi \approx 65.45\%$.  Importantly, rapid increases in deposit volume that coincide with a sudden decrease in impurity scores may correlate with security incidents. Specifically, attackers initially seed an address with minor deposits from \ac{TC} and later use the same channel to launder significant amounts of stolen crypto assets back into \ac{TC}. This pattern aligns closely with 29 cases documented in Section~\ref{sec:limited_restriction_to_security_incidents}.

\textbf{Private Proofs of Innocence in Railgun.} Railgun~\cite{railgunwiki} is an Ethereum-based privacy solution that features a mechanism known as \emph{Private Proofs of Innocence}. Deposits into newly created Railgun shielding tokens must wait for one hour to successfully pass an innocence proof before performing any interaction beyond an immediate refund to the originating address. If this proof fails, the funds are automatically returned to their source. 

We observe the evolving adoption of this mechanism since its activation in November 2023~\cite{popov2024blockchain}. Interestingly, the first detected \ac{TC}-related refund occurred much later, in June 2024, involving account \href{https://etherscan.io/address/0x80df810ed14feb4bbc6ab78a16dd9b0388a8b1b0}{\texttt{0x80Df...B1b0}} at Block 20178593. Prior to this event, we identify transfers totaling 15,780.31 ETH, with impurity score $\varphi \approx 99.35\%$, directly from \ac{TC} into Railgun that did not trigger innocence-proof mechanisms. This finding indicates that Railgun initially allows direct inflows from TC without imposing additional restrictions. Following this specific incident, the Railgun protocol begins to reject funds derived from \ac{TC}. Nonetheless, there are over 2,320.55 ETH deposited into Railgun after the first refund, but among them only around 506.75 ETH (21.84\%) are refunded. This indicates that this mechanism is not effective and users can still successfully circumvent this restriction.

\textbf{Other Privacy-Preserving Tools.} Umbra~\cite{DBLP:conf/www/KovacsS24} leverages stealth addresses to conceal recipient identities. Our analysis identifies 6,286.13 ETH transferred from \ac{TC} into Umbra, carrying a notable impurity score $\varphi \approx 93.07\%$, underscoring its steady adoption. In contrast, Aztec Connect, previously a prominent Ethereum-based privacy solution with inflows of 2,244.93 ETH at impurity score $\varphi \approx 95.71\%$, ceased operations on March 31, 2023~\cite{aztec2023sunsetting}, consequently diminishing its latest relevance. Similarly, another deprecated tool is the Secret Network: Bridge, which received 681.16 ETH with impurity score $\varphi \approx 98.71\%$. Its successor, Secret Tunnel powered by Axelar~\cite{secrettunnel}, continues providing privacy within Secret Network, keeping on-chain transactions and balances fully confidential.

\section{Supplementary Experiment Results}
\label{appendix:supplement}

\begin{table}[h]
\caption{A Typical Test-Depositing Sequence from Transaction History of Address \href{https://etherscan.io/address/0x0e0e9a91a54155bf4b66d66257fbdc8499913143}{\texttt{0x0E0E...3143}}, who contentiously deposited into services with censorship mechanisms (i.e., KuCoin, a \ac{KYC}-compliant centralized exchange).}
\label{tab:test_depositing} 
\centering
\scriptsize
\setlength{\tabcolsep}{3pt}
\resizebox{\columnwidth}{!}{%
\begin{tabular}{llcrrr}
\toprule
Date Time (UTC) & From & Dir & To & Amount (ETH) & Txn Fee (ETH) \\
\midrule
2023-01-13 03:06:47 & \texttt{0x0E0E...3143} & OUT & KuCoin                 & 11.99969862 & 0.00033165 \\
2023-01-13 02:28:59 & \texttt{0x4BaC...671F} & IN  & \texttt{0x0E0E...3143} & 12.00000000 & 0.00072839 \\
\midrule
2023-01-10 12:06:35 & \texttt{0x0E0E...3143} & OUT & KuCoin                 & 11.99959166 & 0.00040296 \\
2023-01-10 11:21:35 & \texttt{0xC728...D5A1} & IN  & \texttt{0x0E0E...3143} & 12.00000000 & 0.00031041 \\
\midrule
2023-01-09 10:07:11 & \texttt{0x0E0E...3143} & OUT & KuCoin                 & 10.59799828 & 0.00034590 \\
2023-01-09 10:01:11 & \texttt{0x2c03...c42D} & IN  & \texttt{0x0E0E...3143} & 10.59838156 & 0.00031670 \\
\midrule
2023-01-08 10:06:47 & \texttt{0x0E0E...3143} & OUT & KuCoin                 & 11.99973839 & 0.00030656 \\
2023-01-08 09:49:11 & \texttt{0x2c03...c42D} & IN  & \texttt{0x0E0E...3143} & 12.00000000 & 0.00037588 \\
\midrule
\rowcolor{mylightyellow} 2023-01-01 14:06:35 & \texttt{0x0E0E...3143} & OUT & KuCoin                 & 0.09968959  & 0.00026543 \\
\rowcolor{mylightyellow} 2023-01-01 13:19:11 & \texttt{0x2C48...0B8B} & IN  & \texttt{0x0E0E...3143} & 0.10000000  & 0.00031183 \\
\bottomrule
\end{tabular}%
}
\end{table}

\begin{table}[b]
\caption{Active Span and Impurity Volume of \ac{CEX}}
\label{tab:cex_volume_and_span}
\centering
\resizebox{\linewidth}{!}{%
\begin{tabular}{lrr}
\toprule
\ac{CEX}    & Active Span (Days)     & Impurity Volume $\mathcal{I}_v$ \\
\midrule
Binance                             & 822 (85.89\%) & 20,097.20 ETH (16.33\%) \\
\rowcolor{mylightyellow} ChangeNOW  & 802 (83.80\%) &  3,942.41 ETH (03.20\%) \\
\rowcolor{mylightyellow} FixedFloat & 702 (73.35\%) & 25,450.88 ETH (20.68\%) \\
\rowcolor{mylightyellow} eXch       & 609 (63.64\%) & 22,018.98 ETH (17.89\%) \\
Coinbase                            & 584 (61.02\%) &    960.88 ETH (00.78\%) \\
KuCoin                              & 581 (60.71\%) &  5,422.31 ETH (04.40\%) \\
MEXC                                & 557 (58.20\%) &  6,504.51 ETH (05.28\%) \\
OKX                                 & 500 (52.25\%) & 15,420.37 ETH (12.53\%) \\
Bybit                               & 430 (44.93\%) &  3,064.73 ETH (02.49\%) \\
SideShift                           & 286 (29.89\%) &  1,452.07 ETH (01.18\%) \\
HTX                                 & 254 (26.54\%) &  5,760.38 ETH (04.68\%) \\
Kraken                              & 214 (22.36\%) &    408.93 ETH (00.33\%) \\
Gate.io                             & 203 (21.21\%) &  4,007.67 ETH (03.26\%) \\
Bitget                              & 115 (12.02\%) &    546.41 ETH (00.44\%) \\
SimpleSwap                          & 114 (11.91\%) &  4,516.20 ETH (03.67\%) \\
TradeOgre                           & 110 (11.49\%) &  3,525.23 ETH (02.86\%) \\
\midrule
Total & 957 \hspace{4pt} (100\%) & 123,099.20 ETH \hspace{4pt} (100\%) \\
\bottomrule
\end{tabular}
}
\end{table}

\end{document}